\documentclass[reprint, amsmath, amssymb, aps, superscriptaddress, nolongbibliography, floatfix]{revtex4-2}
\usepackage{amsfonts
}
\usepackage{amsmath}
\usepackage{amssymb}
\usepackage{graphicx}
\usepackage{color}
\usepackage{braket}
\usepackage[utf8]{inputenc}
\usepackage{hyperref} 
\usepackage{ytableau, comment}
\usepackage{qcircuit}
\usepackage{booktabs}
\usepackage{float}

\def \be {\begin{equation}}
\def \ee {\end{equation}}
\def \ba {\begin{aligned}}
\def \ea {\end{aligned}}
\def \bea {\begin{eqnarray}}
\def \eea {\end{eqnarray}}

\begin{document}

\title{Effective Noise Mitigation via Quantum Circuit Learning in Quantum Simulation of Integrable Spin Chains}

\author{Wenlong Zhao}\altaffiliation{First authors}
\affiliation{School of Physical Science and Technology, Soochow University, Suzhou 215006, China}
\author{Yimeng Zhang}\altaffiliation{First authors}
\affiliation{School of Physical Science and Technology, Soochow University, Suzhou 215006, China}
\author{Yan Guo}\altaffiliation{First authors}
\affiliation{School of Physical Science and Technology, Soochow University, Suzhou 215006, China}
\author{Yufan Cui}
\affiliation{School of Physical Science and Technology, Soochow University, Suzhou 215006, China}
\author{Zhuohang Wang}
\affiliation{School of Physical Science and Technology, Soochow University, Suzhou 215006, China}
\author{Rui-Dong Zhu}\altaffiliation{Corresponding author}
\affiliation{School of Physical Science and Technology, Soochow University, Suzhou 215006, China}
\affiliation{Institute for Advanced Study, Soochow University, Suzhou 215006, China}
\affiliation{Jiangsu Key Laboratory of Frontier Material Physics and Devices,\\ Soochow University, Suzhou 215006, China}

\begin{abstract}
We propose a noise-mitigation quantum simulation strategy for near-term quantum devices based on Quantum Circuit Learning (QCL), which is in particular effective for integrable quantum spin chains. The method trains a shallow variational circuit to approximate a deeper time-evolution circuit by learning the conserved charges and only a small amount of dynamical information in the system. Under realistic noise models, the learned circuit maintains both conserved quantities and dynamical observables significantly closer to their true values than the noisy simulation of the original circuit. We demonstrate, on small-scale prototypes, that QCL can act as an effective, physics-informed error mitigation strategy, producing shorter, more robust circuits without exponential sampling overhead.

 \end{abstract}
\maketitle

\section{Introduction}

Quantum many-body problems, particularly in strongly coupled and highly correlated regimes, have long been recognized as fundamental challenges in physics, epitomized by ``More is Different''. For computing energy spectra and other equilibrium properties, physicists have developed a variety of classical numerical methods. However, for simulating the real-time dynamical evolution of such systems, quantum computers hold great promise. Yet, current quantum hardware operates in the noisy intermediate-scale quantum (NISQ) era, where gate infidelities, decoherence, and readout errors severely constrain achievable circuit depth and computational fidelity. The results obtained from running quantum circuits on a real device often exhibit substantial deviations from their theoretical predictions. The development of effective error-mitigation strategies is therefore critical to enabling practical quantum simulation.

Quantum circuit learning (QCL) proposed in \cite{Mitarai:2018voy} has emerged as a powerful quantum machine learning framework wherein parameterized quantum circuits are trained to fit target functions or solve classification problems. In a typical QCL setup, classical input data is first embedded into a quantum state, after which a sequence of parameterized quantum gates—often composed of rotational and entangling layers—processes the state. The output is obtained by measuring a designated observable, and its expectation value is interpreted as the prediction of the model. Beyond data fitting problems, QCL has also been successfully applied to classification tasks\cite{Havlicek2019,Li2021,Schnabel2025,Suzuki2024}, where the measured expectation value or a set of such values is mapped to discrete class labels through a post-processing function or a classical decision layer. Compared to classical neural networks, QCL leverages inherent quantum properties such as superposition and entanglement, which can offer expressive functional representations within relatively compact circuit \cite{Heimann:2022ytr}. 

In this work we therefore explore a new route to quantum error mitigation that is particularly tailored to integrable models. We leverage the extensive set of conserved charges in integrable models to train an effective, low-depth quantum circuit that reproduces the action of a much deeper time-evolution circuit on a randomly-chosen manifold of input states. We expect such an effective but shallower quantum circuit to be less influenced by the quantum error in realistic noisy simulations. The conserved quantities in the integrable model provide a dense and robust training data for the effective quantum circuit, while at the same time acting as stringent benchmarks of noise-induced deviations, as proposed in \cite{Maruyoshi:2022jnr}. This “physics-informed circuit compression” is performed classically in this work, but once learned, the shallow effective circuit can be executed on hardware a single circuit instance per inference point, without the multiple noise-scaled runs required by schemes such as zero-noise extrapolation or the sampling overhead of probabilistic error cancellation. 


This article is organized as follows. In Sec. \ref{s:int}, we introduce the necessary background on the integrable spin chain models and their quantum simulations. Sec. \ref{s:QCL} reviews how the QCL works and then details our methodology, explaining how the conserved charges of the model are integrated into the training process and describing the variational circuit ansatz. Our main numerical results are presented in Sec. \ref{s:res}, beginning with the learning outcomes and noise mitigation performance for a two-qubit chain across four common noise channels, and then extending to a three-qubit system to illustrate the applicability of our approach beyond the minimal two-qubit prototype. Finally, Sec. \ref{s:con} summarizes the findings, discusses the implications and advantages of our QCL-based approach compared to conventional error mitigation techniques, and outlines several promising directions for future work.

\section{Integrable Spin Chains and Conserved Charges}\label{s:int}

In the quantum simulation of integrable spin chains, one may employ a Trotterized approximation $U(\delta)$ of the continuum time evolution operator and realize it with quantum gates \cite{Lloyd:1996aai}. $\delta$ represents an infinitesimal evolution time (multiplied by the coupling $J$), and in the continuum limit, 
\begin{equation}
    \lim_{d\to \infty}U(\delta=-Jt/d)^d={\rm e}^{-iHt}.
\end{equation}
In this work, we focus on the quantum simulation of XXX Heisenberg spin chain described by the Hamiltonian, 
\begin{equation}
    H=\frac{J}{2}\sum_{j=1}^LX_{j}X_{j+1}+Y_{j}Y_{j+1}+Z_{j}Z_{j+1}.\label{XXX}
\end{equation}
This model is exactly solvable and was first solved by H. Bethe in almost a century ago \cite{Bethe:1931hc}. Its solvablity comes from the existence of the same number of the conserved charges as the d.o.f.'s in the system. In the algebraic Bethe ansatz approach (see lecture notes \cite{Faddeev:1994nk,Slavnov:2018kfx}), these conserved charges can be constructed from the R-matrix, a solution to the Yang-Baxter equation, and the transfer matrix defined from the R-matrix. 
For spin-$1/2$ XXX model, the (braided) R-matrix is given by 
\begin{equation}
\check{R}(u) = \frac{1 + iu P}{1 + iu},
\end{equation}
where $P=\frac{1}{2}\left(1+X\otimes X+Y\otimes Y+Z \otimes Z\right)$ is the permutation operator. When the spin chain length $L$ is even, the discrete time evolution operator $U(\delta)$ for a single time step can be expressed as \cite{2018PhRvL.121c0606V}
\begin{equation}
U(\delta) = \prod_{j=1}^{L/2} \check{R}_{2j-1,2j}(\delta) \prod_{j=1}^{L/2} \check{R}_{2j,2j+1}(\delta),\label{time-evo}
\end{equation}
and it is related to the transfer matrix $T(u)$,
\begin{equation}
    T(u)={\rm tr}_0\left(\prod_{1\leq j\leq L}^{\leftarrow}R_{0j}\left(u-\frac{(-1)^j}{2}\delta\right)\right),
\end{equation}
by the following relation, 
\begin{equation}
    U(\delta)=T^{-1}(-\delta/2)T(\delta/2),
\end{equation}
where $R(u):=P\check{R}(u)$. The concrete realization of this time evolution operator in terms of quantum gates can be found in e.g. \cite{Maruyoshi:2022jnr}, and in general the simulation circuit is very long even for a few-body system. From the commutativity of the transfer matrix $[T(u),T(v)]=0$ for $^\forall u,v$, one can immediately see that the transfer matrix keeps the same along the time evolution of the system. This gives a family of conserved charges in the integrable system via 
\begin{equation}
    Q^\pm_n:=\left.\frac{{\rm d}^n}{{\rm d}u^n}\log T(u)\right|_{u=\pm\frac{\delta}{2}}.
\end{equation}

It was proposed in \cite{Maruyoshi:2022jnr} that one can use the above conserved charges as benchmarks to evaluate the influence of noise in quantum simulation. In this work, we adopt this idea to estimate the noise reduction effect in our quantum circuit. Let us give the concrete expression for $Q_1^\pm$ here, 
\begin{align}
    Q^+_1=\sum_{n=1}^{L/2}\frac{i}{2(1+\delta^2)}q^{(1)}_{2n-2,2n-1,2n}(+,\delta),\\
    Q^-_1=\sum_{n=1}^{L/2}\frac{i}{2(1+\delta^2)}q^{(1)}_{2n-1,2n,2n+1}(-,\delta),
\end{align}
where
\begin{align}
q^{(1)}_{i,j,k}(\sigma,\delta) := \vec{\sigma}_i\cdot\vec{\sigma}_j + \vec{\sigma}_j\cdot\vec{\sigma}_k+\delta^2\vec{\sigma}_k\cdot\vec{\sigma}_i\cr - \sigma\delta\,\vec{\sigma}_i\cdot(\vec{\sigma}_j\times\vec{\sigma}_k),
\end{align}
and we denoted $\vec{\sigma}_j=(X_j,Y_j,Z_j)$ for simplicity. 

In addition, the system \eqref{XXX} has an SU(2) symmetry generated by the total spin operator $\vec{S}:=\sum_{i=1}^L\vec{\sigma}_i$, and these operators give three extra conserved charges. 

In this article, we mainly use the cases of $L=2$ and $L=3$ to demonstrate our idea. In the simplest case of $L=2$, there are only four d.o.f.'s in the system, and it is enough to use $\vec{S}=(X_{\rm tot},Y_{\rm tot},Z_{\rm tot})$ and the Hamiltonian ${\cal H}=2\vec{\sigma}_1\cdot \vec{\sigma}_{2}$ to solve the model. When we take $L=3$, the time evolution operator \eqref{time-evo} cannot be defined, and we use instead 
\begin{equation}
    U_3(\delta)=\check{R}_{12}(\delta)\check{R}_{23}(\delta)\check{R}_{31}(\delta),\label{U3}
\end{equation}
to simulate the dynamics of a near-integrable model. One can identify four rigorously conserved charges, $\vec{S}$ and ${\cal H}=\sum_{i=1}^L\vec{\sigma}_i\cdot \vec{\sigma}_{i+1}$ in this model, and we also found four approximately conserved charges ${\cal C}_1^\pm:=q^{(1)}_{123}(\pm,\delta)$, ${\cal C}_2^\pm=q^{(1)}_{312}(\pm,\delta)+q^{(1)}_{231}(\pm,\delta)$. They are almost invariant under the time evolution up to ${\cal O}(\delta^2)$ corrections, 
\begin{equation}
    \left(U^\dagger_3\right)^n{\cal C}_i^\pm U_3^n-{\cal C}_i^\pm={\cal O}(\delta^2).
\end{equation}
By taking $\delta$ sufficiently small, these quantities remain approximately conserved and provide useful constraints for analyzing the dynamics.

\section{Quantum Circuit Learning for Spin Chains} \label{s:QCL}

The quantum circuit learning proposed in \cite{Mitarai:2018voy} is a machine learning method based on quantum circuits instead of neural networks.  Its core idea involves encoding classical data into quantum states, which are then processed by a parameterized quantum circuit composed of tunable quantum gates. The output is typically obtained by measuring the expectation value of an observable $M$, e.g., a Pauli operator, with respect to the final quantum state. Formally, for input data $x$ and parameters $\theta$, the output can be expressed as, 
\begin{equation}
f_M(x;{\boldsymbol{\theta}})={}^{\otimes L}\bra{0}V^\dagger(x,\{\theta_i\})MV(x,\{\theta_i\})\ket{0}^{\otimes L},\label{f-loss}
\end{equation}
where $V(x,\{\theta_i\})$ represents the unitary operator that incorporates both the data encoding and the parameterized quantum gates. The parameters $\{\theta_i\}$ of the circuit are trained either using classical optimizers like gradient descent or a pure quantum measurement to determine the gradient. According to the discussion in \cite{Heimann:2022ytr}, \eqref{f-loss} is essentially a truncated, real-valued Fourier series, and one can use it to fit any function. 

Compared with classical neural networks, QCL offers potential advantages: the superposition and entanglement inherent in quantum states may enable more compact data representations and more efficient feature extraction, while the inherent parallelism of quantum operations could provide speed-ups for certain high-dimensional problems. However, the practical performance of QCL is currently constrained by hardware limitations such as qubit count, circuit depth, and noise, making it an active and evolving area of research.

In our work, on the other hand, we utilize the QCL to learn quantum states in integrable models. The candidate state is prepared in a similar quantum circuit. First, we choose a family of input states parameterized by $x \in [-1, 1]$ generated from the all-up state $\ket{0}^{\otimes L}$ via a unitary transformation,
\begin{equation}
\begin{split}
V_{\text{in}}(x) = & \bigotimes_{j=1}^{L/2} \left[ R_j^Z(\cos^{-1}x^2) R_j^Y(\sin^{-1}x) \right] \\
& \otimes \left[ R_{j+1}^Z(\cos^{-1}x^2) R_{j+1}^X(\sin^{-1}x) \right],
\end{split}
\end{equation}
where $R_j^A(\theta) = \exp(-i\theta A_j/2)$ for $A=X,Y,Z$. This will be referred to as the initial state $\ket{\psi_{\rm in}(x)}$ in our quantum simulation, and it is chosen not to be an eigenstate of the XXX Hamiltonian. In the learning circuit, we apply the following parameterized quantum gates $W_I$,
\begin{align}
\Qcircuit @C=0.3em @R=0.2em {
& \multigate{3}{e^{iH_dT}} &  \gate{R_x(\theta_{3LI+1})} & \gate{R_z(\theta_{3LI+2})} &  \gate{R_x(\theta_{3LI+3})} &\qw \\
& \ghost{e^{iH_dT}}& \gate{R_x(\theta_{3LI+4})} & \gate{R_z(\theta_{3LI+5})} &  \gate{R_x(\theta_{3qI+6})} \qw \\
& \ghost{e^{iH_dT}} & \qw & \vdots & & \qw \\
& \ghost{e^{iH_dT}} & \gate{R_x(\theta_{3L(I+1)-2})} & \gate{R_z(\theta_{3L(I+1)-1})} &  \gate{R_x(\theta_{3L(I+1)})} & \qw}
\end{align}
for $D$ times to generate the output state. 
\begin{equation}
    \ket{\psi_{\rm out}(x,\{\theta_i\})}=\prod_{I=1}^DW_I\left(\{\theta_j\}_{j=3LI+1}^{3L(I+1)}\right)\ket{\psi_{\rm in}(x)}.\label{goal}
\end{equation}
$W_I$ consists of a time evolution gate $e^{-i\mathcal{H}_dT}$ with $H_d$ a different (and simpler) Hamiltonian, and a sequence of single-qubit rotations, $R_j^X(\theta_{3kL+j})R_j^Z(\theta_{3kL+j+1})R_j^X(\theta_{3kL+j+2})$, for each qubit $j$. 

Our goal is to reproduce the time evolution operator $U(\delta)^d$ by $W_I$'s with proper variational parameters $\{\theta^{(d)}_i\}$. That is in the Hilbert subspace of $\ket{\psi_{\rm in}(x)}$, ${\cal H}_{\rm in}$, 
\begin{equation}
    \left.U(\delta)^d\right|_{{\cal H}_{\rm in}}\approx \left.\prod_{I=1}^DW_I\left(\{\theta^{(d)}_j\}\right)\right|_{{\cal H}_{\rm in}}.\label{eq:approx}
\end{equation}
This is similar to the quantum compiling and unitary learning techniques in the literature \cite{Khatri:2018mjs,Coles:2018vmk}, but as will be discussed later, a different training progress will be taken by fully leveraging the power of conserved charges in the system. 

The advantage of doing so is that by choosing a simple enough $H_d$, one can effectively realize the time evolution for $\ket{\psi_{\rm in}(x)}$ with a shorter circuit, and we expect that such a circuit will reduce the influence of noise in a real quantum simulation on noisy quantum devices. In our work, we follow the original framework of \cite{Mitarai:2018voy} to use an all-to-all Ising Hamiltonian, 
\begin{equation}
    H_d=\sum_{j\neq k}a_{j,k}Z_jZ_k,\label{H-train}
\end{equation}
to construct the learning circuit. $a_{j,k}$ are random couplings, and the time evolution operator $e^{iH_d T}$ can be realized with a series of the following two-qubit gates ${\rm e}^{ia_{j,k}Z_jZ_k}$, 
\begin{align}
\Qcircuit @C=1em @R=.7em {
\lstick{j} &\qw & \ctrl{2} & \qw & \ctrl{2} & \qw\\
\lstick{\vdots} & &  & & & \\
\lstick{k} &\qw & \targ & \gate{R_z(-2a_{j,k})} & \targ & \qw}
\end{align}
In particular, $D$ can be fixed to a relatively small number, e.g. $D=3$ or $D=4$, which means that a short circuit can approximate a very long simulation circuit even for large $N$. We thus expect that the learning circuit in real simulation will not be affected by the noise as much as the original circuit. 

To train the parameterized circuit for \eqref{goal}, we use the conserved charges, which can be read from $\ket{\psi_{\rm in}(x)}$, and a dynamical observable $\langle Z_1\rangle$ depending on $d$ as input data, which needs to be computed directly or with the Bethe ansatz method. As discussed in \cite{Mitarai:2018voy}, this enables us to perform a pure quantum training of the circuit, where the gradient for each $\theta_i$ can be computed from shifted measurements, but in this work, we performed a classical simulation to find the optimized variational parameters. 

More concretely, we divide the interval $x \in [-1, 1]$ into $M=200$ points $\{x_i\}$, and prepare the training datasets $\{(x_i, y_{i,{\rm train}}^{O})\}_{i=1}^M$, where $y_{i,{\rm train}}^{O}=\bra{\psi_{\rm in}(x_i)} O\ket{\psi_{\rm in}(x_i)}$ is the expectation value for $O$ taking value in the set of conserved charges or $Z_1$. The loss function for the quantum circuit learning is then set to be 
\begin{equation}
\begin{split}
\mathcal{L}(\boldsymbol{\theta}, a) = &\; \alpha_1 \frac{1}{M}\sum_i \left( a f_{Z_1}(x_i,{\boldsymbol{\theta}}) - y_{i,\text{train}}^{Z_1} \right)^2 \\
&+ \alpha_2 \frac{1}{M}\sum_i \left( a f_{Z_{\rm tot}}(x_i,{\boldsymbol{\theta}}) - y_{i,\text{train}}^{Z_{\rm tot}} \right)^2 \\
&+ \alpha_3 \frac{1}{M}\sum_i \left( a f_{X_{\rm tot}}(x_i,{\boldsymbol{\theta}}) - y_{i,\text{train}}^{X_{\rm tot}} \right)^2 \\
&+ \alpha_4 \frac{1}{M}\sum_i \left( a f_{Y_{\rm tot}}(x_i,{\boldsymbol{\theta}}) - y_{i,\text{train}}^{Y_{\rm tot}} \right)^2 \\
&+ \alpha_5 \frac{1}{M}\sum_i \left( a f_{H}(x_i,{\boldsymbol{\theta}}) - y_{i,\text{train}}^{H} \right)^2 .
\end{split}
\end{equation}
The coefficients $\alpha_k$ are hyperparameters that balance contributions from different observables. The trainable parameters consist of the variational circuit parameters $\boldsymbol{\theta}=\{\theta_i\}$ and a global scaling factor $a$. We note that the inclusion of an unnecessary parameter $a$ aims to accelerate the learning process, for consistency, its final optimized value is expected to remain close to $1$. As we will see later, a typical value for the optimized $a$ stays at $|a-1|\lesssim0.05$.

We employ the Nelder-Mead simplex algorithm for parameter optimization, motivated by its robustness in the presence of noisy objective landscapes. Variational parameters $\{\theta_i\}$ are initialized uniformly at random from \([0, 2\pi]\), and the global scaling factor \(a\) is initially set to unity. In each training epoch, the loss function is evaluated by executing the quantum circuit on all training inputs. The parameters are then updated according to the Nelder-Mead operations of reflection, expansion, contraction, and shrinkage. The optimization terminates when the change in the loss function falls below a predefined tolerance or when the maximum number of iterations is reached. We will refer to the trained variational circuit as the {\it learned circuit} or {\it QCL circuit} throughout this article. 

\begin{figure*}[!htb]
  \centering
    \includegraphics[width=\textwidth]{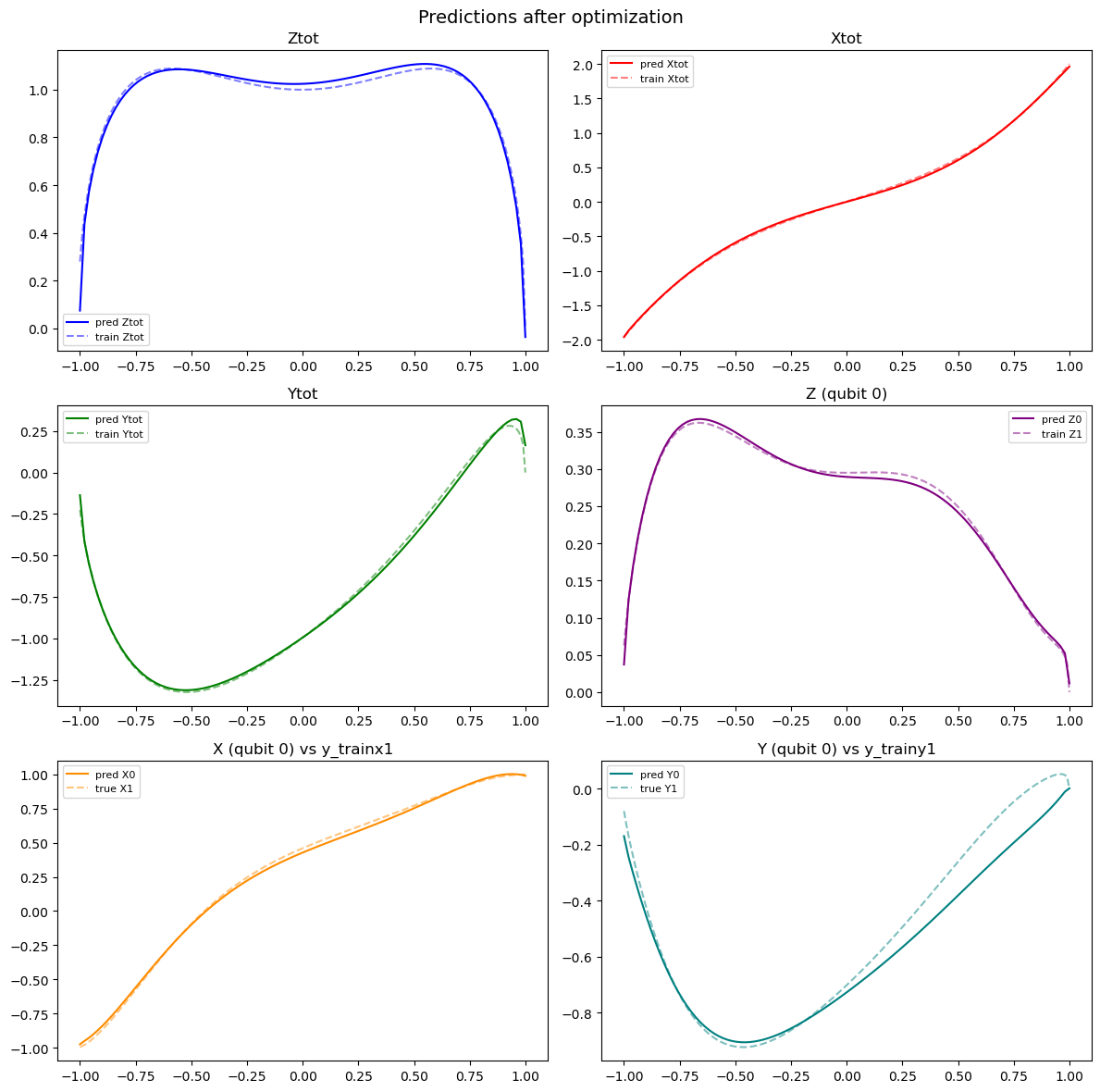}
    \caption{Predictions from learned circuit (solid lines) compared with the original data for $L=2$ and $d=15$ (dashed lines). In particular, the data for $\langle X_1\rangle$ and $\langle Y_1\rangle$ were not taught to the circuit ({\it n.b.} in the quantum simulation, the first qubit is usually called qubit 0). The scaling factor $a$ is optimized to $a_{\rm opt}=1.014819$. }
    \label{fig:learn}
\end{figure*}

For example, in the case of $L=2$ $d=15$, we used a $D=3$ circuit to learn the evolved quantum state. Figure \ref{fig:learn} shows the comparison between the true expectation values computed from the original quantum circuit (dashed lines) and the predictions from the learned circuit (solid lines). We remark that even though the quantum circuit was not provided with any information about the observables $\langle X_1\rangle $ and $\langle Y_1\rangle$, it predicted their values rather accurately. This capability is attributed to the sufficient information from the conserved quantities of the integrable model that was fed into the training process.

The learned circuit is then passed to a noisy quantum simulation, where we test our idea to see whether it behaves better under the influence of noise than the original quantum circuit.

\section{Results: Noisy Simulation of Learned Circuit}\label{s:res}

In this section, we present the results of noisy simulation for the quantum circuit trained by the QCL method and compare its behavior with that of the original quantum simulation circuit under the same noise to show the noise mitigation. To simulate a realistic quantum device, we adopt the following four commonly used noise models in our noisy simulation:
\begin{itemize}
    \item Bit-flip noise: Each gate operation incurs a Pauli $X$ error with probability $p$. Its Kraus representation is given by $E_0 = \sqrt{1-p}I$, $E_1 = \sqrt{p}X$.
    \item Depolarizing noise: With probability $p$, the quantum state is replaced by the maximally mixed state. For a single qubit, the model is described by the Kraus operators: $E_0 = \sqrt{1-3p/4}I$, $E_1 = \sqrt{p/4}X$, $E_2 = \sqrt{p/4}Y$, $E_3 = \sqrt{p/4}Z$.
    \item Amplitude damping: This channel models energy relaxation with probability $p$, described by $E_0 = \begin{pmatrix}1 & 0 \\ 0 & \sqrt{1-p}\end{pmatrix}$, $E_1 = \begin{pmatrix}0 & \sqrt{p} \\ 0 & 0\end{pmatrix}$.
    \item Phase damping: This model describes dephasing with probability $p$, and is represented by Kraus operators $E_0 = \sqrt{1-p/2}I$, $E_1 = \sqrt{p/2}Z$.
\end{itemize}
In this work, we primarily employ an $L=2$ prototype to demonstrate the noise mitigation effect, but the underlying methodology is readily generalizable to systems with more qubits (as will be briefly discussed later) and to the simulation of other integrable models.

\subsection{Learning results}

\begin{figure*}[!htb]
\centering
\includegraphics[width=0.8\textwidth]{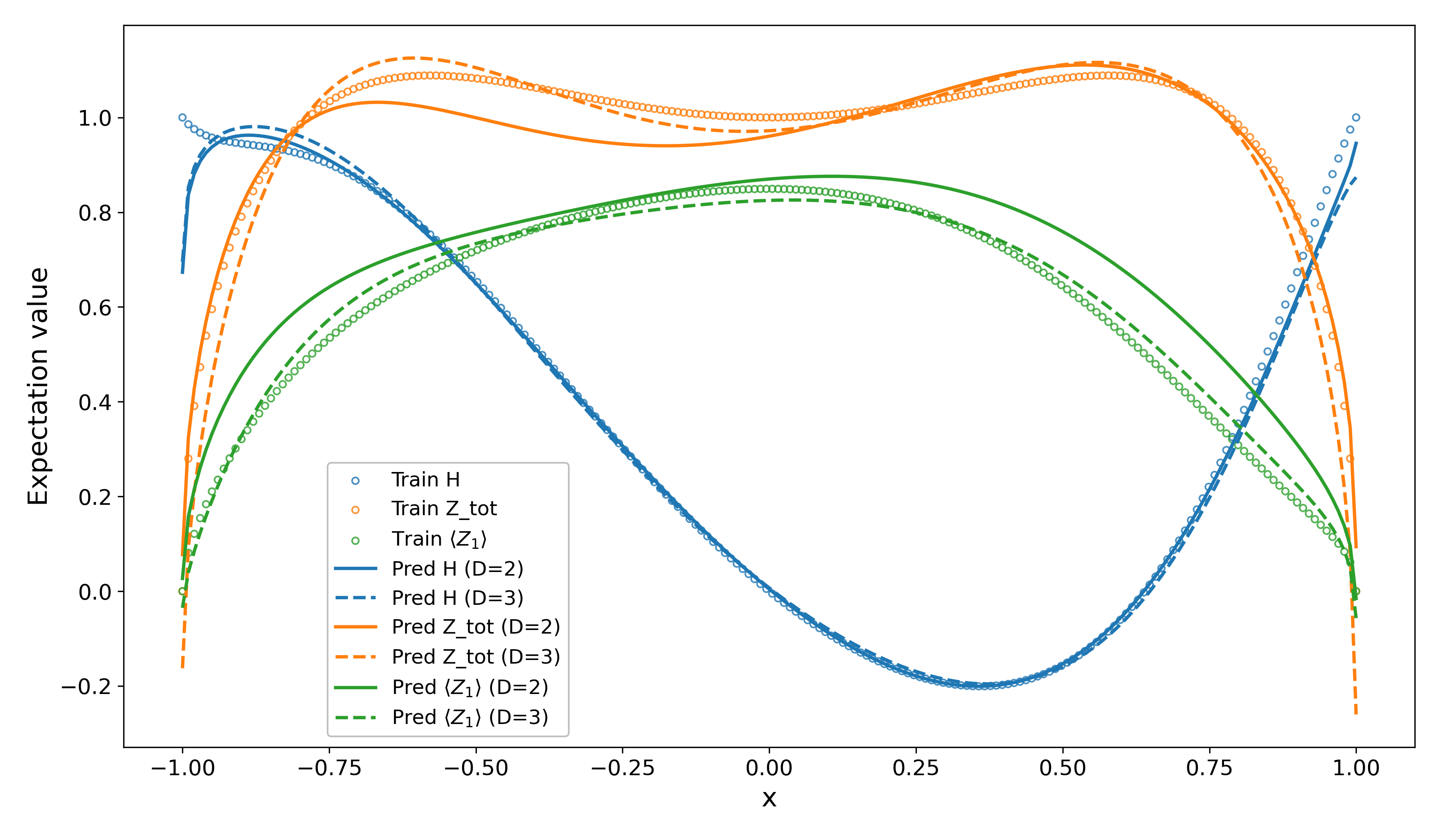}
\caption{Comparison of predictions for the training data, $H$, $Z_{\rm tot}$, and the dynamical observable $\langle Z_1 \rangle$ at $d=4$ using QCL with circuit depths $D=2$ (solid lines) and $D=3$ (dashed lines). Both setups achieve relatively good agreement with the training data, and give similar predictions. }
\label{fig:all_in_one_comparison}
\end{figure*}

\begin{figure*}[!htb]
\centering
\includegraphics[width=0.8\textwidth]{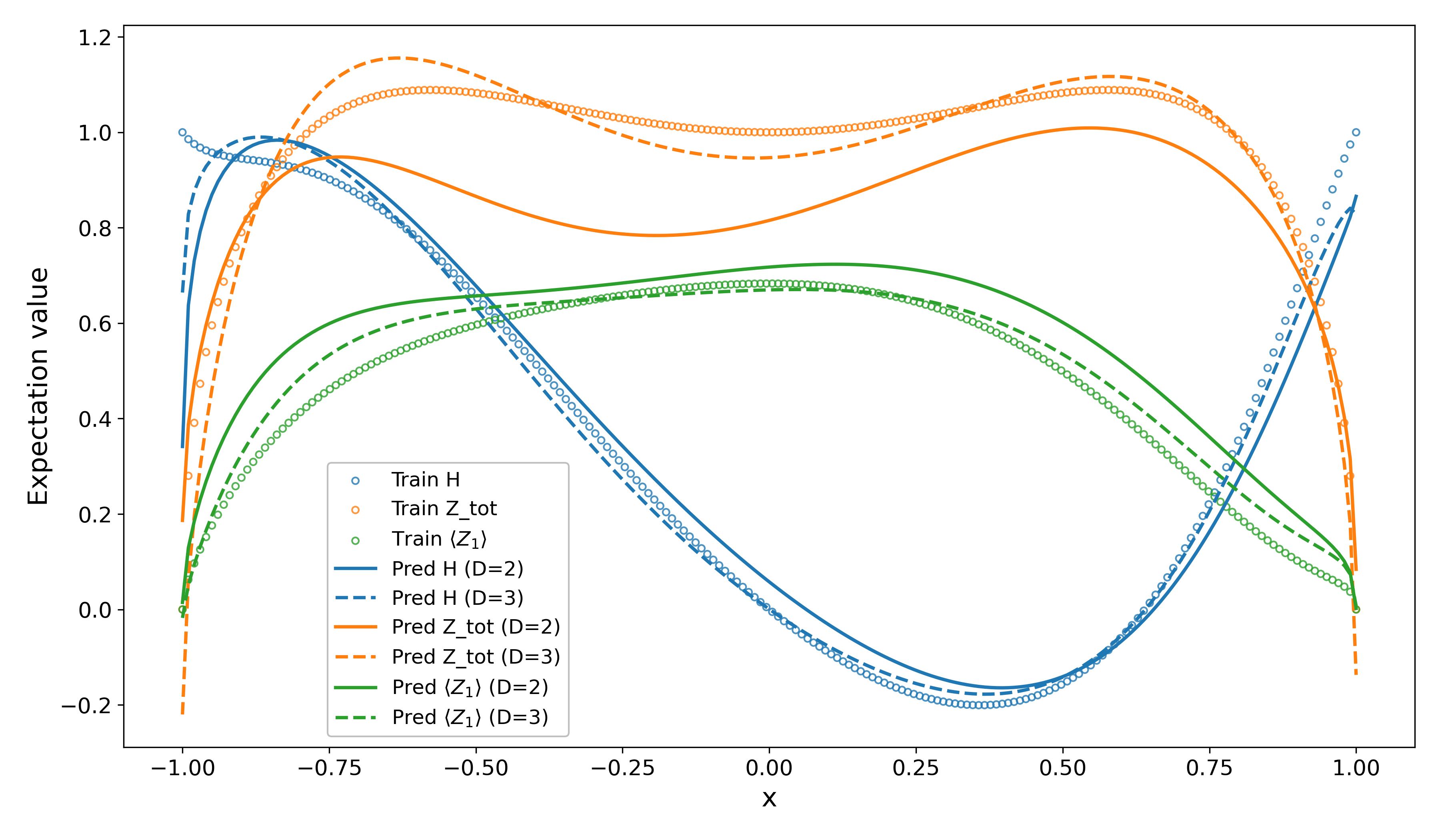}
\caption{Comparison of predictions for the training data, $H$, $Z_{\rm tot}$, and the dynamical observable $\langle Z_1 \rangle$ at $d=6$ using QCL with circuit depths $D=2$ and $D=3$. The predictions of the $D=3$ QCL circuit (dashed lines) are clearly better than those of $D=2$ circuit, especially for $\langle Z_{\rm tot}\rangle$. }
\label{fig:all_in_one_comparison-6}
\end{figure*}

One can choose the number of free parameters in the learning circuit by tuning $D$. For $L=2$, we performed training respectively with $D=2$, $3$, and $4$, and selected the best result as evaluated by the loss function, where the parameters were set as $\alpha_1=2$, $\alpha_2=\alpha_3=\alpha_4=1$, and $\alpha_5$ flexibly chosen between $2$ and $4$ according to the variation of training data. The training outcome proved sensitive to randomly sampled initial parameters, as the optimization could become trapped in local minima. The loss function values after multiple optimization trials are typically reduced to $\mathcal{O}(10^{-3})$, compared to initial values of $\mathcal{O}(10^0)$, confirming successful training across all configurations. We adopted $D=2$ for the time evolution step $d=\{1,2,3,4\}$ and $D=4$ for $d={6,8,10,12,15}$ in the noisy simulations. We remark that when $d$ is small, the time evolution operator $U(\delta)^d$ is close to the identity and we only need a small number of rotational gates to generate the same effect. For example, at $d=4$, the predictions by learned circuit of depth $D=3$ only improve by a little compared to those of $D=2$ (see Figure \ref{fig:all_in_one_comparison}), so we pick the $D=2$ circuit for subsequent simulations. It is natural to expect that a deeper learning circuit with larger $D$ is necessary to approximate a longer time evolution. Indeed, even for $d=6$, the QCL circuit with $D=3$ yields significantly better predictions compared to that with $D=2$ (see Figure \ref{fig:all_in_one_comparison-6}).

We added a redundant variational parameter $a$ to improve and accelerate the optimization process. The approximation \eqref{eq:approx} only works well with a low final loss function and $a$ close to $1$. 
 Table~\ref{tab:a_parameters} lists the optimized values for $a$ in the learning circuit adopted to noisy simulation. 

\begin{table}[H]
\centering
\begin{tabular}{cccccc}
\toprule
$D$ & 2 & 2 & 2 & 2 & 3 \\
$d$ & 1 & 2 & 3 & 4 & 6 \\
$a$ & 0.987135 & 0.974771 & 0.966193 & 0.985416 & 0.995722 \\
\bottomrule
\end{tabular}
\qquad
\begin{tabular}{cccccc}
\toprule
$D$ & 3 & 4 & 4 & 4 & 4 \\
$d$ & 10 & 6 & 8 & 12 & 15 \\
$a$ & 1.014819 & 1.004481 & 1.060125 & 0.979368 & 1.006795\\
\bottomrule
\end{tabular}
\par\vspace{2mm}
\caption{Optimized scaling factors $a$ for various $(D, d)$ used in noisy simulation.}
\label{tab:a_parameters}
\end{table}

In addition to the final loss value and scaling factor $a$, one can directly use the fidelity to evaluate how good the learning outcome is. We define the overlap between the output state of the learned circuit and the true state obtained from the time evolution $U(\delta)^d\ket{\psi_{in}(x)}$ as 
\begin{equation}
    F(x;d,D):=\left|\bra{\psi_{out}(x,\{\theta^{\rm opt}_i\})}U(\delta)^d\ket{\psi_{in}(x)}\right|.
\end{equation}
The values of overlap in our learned circuits are all above $0.9$, showing a relatively good approximation achieved by the QCL. For example, for a randomly chosen value of $x$, $F(0.1,d=4,D=2)=0.923$, $F(0.3,d=10,D=3)=0.998$, $F(-0.1,d=15,D=3)=0.927$ \footnote{For $d=15$, the learning circuit with $D=4$ gives better results than $D=3$, so $D=4$ circuit was used in the noisy simulation. The learning result for $d=15$ and $D=3$ appears in Figure \ref{fig:learn}, and we include it for comparison of the overlap. }, and $F(-0.3,d=15,D=4)=0.985$. The overlap is almost independent of $x$. 

\subsection{Noise mitigation effect}

To examine our idea that QCL can produce an effective, shorter-depth simulation circuit and thereby mitigate noise impact, we conducted noisy simulations under the four previously mentioned noise models. The noise probabilities were set to $p=1\%$ for depolarizing noise, amplitude damping, and  phase damping models, and $p=0.5\%$ for bit-flipping noise, representing moderate yet realistic noise levels. 

\begin{figure}[!htb]
    \centering
    \includegraphics[width=0.48\textwidth]{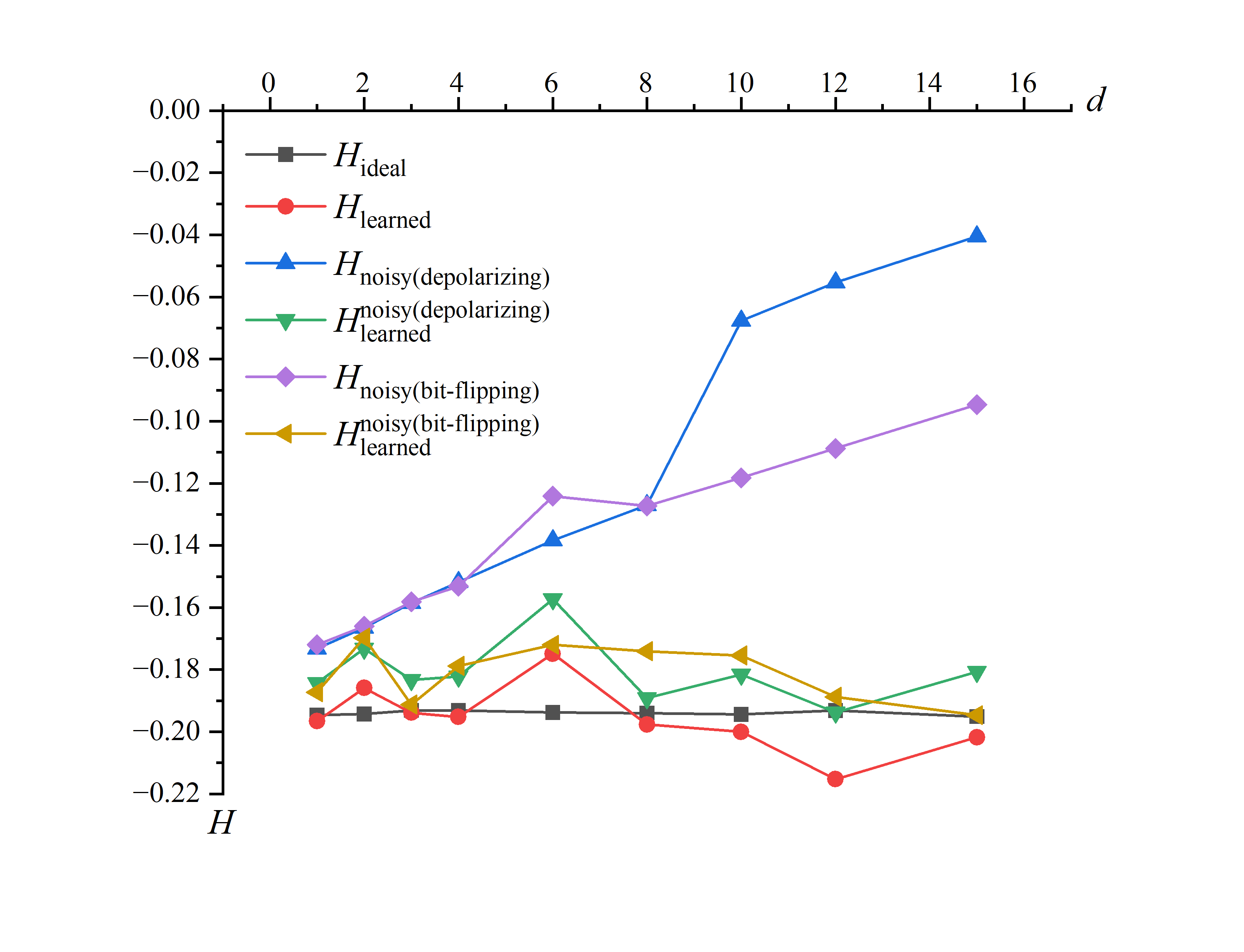}
    \caption{Conserved Hamiltonian $H$ at $x=0.3$ under bit‑flipping ($p=0.5\%$) and depolarizing ($p=1\%$) noise. The learned circuit closely follows the ideal curve, while the unmitigated simulation quickly deviates as the time‑evolution depth $d$ increases.}
    \label{fig:H_bit_dep_x0_3}
\end{figure}

\begin{figure}[!htb]
    \centering
    \includegraphics[width=0.48\textwidth]{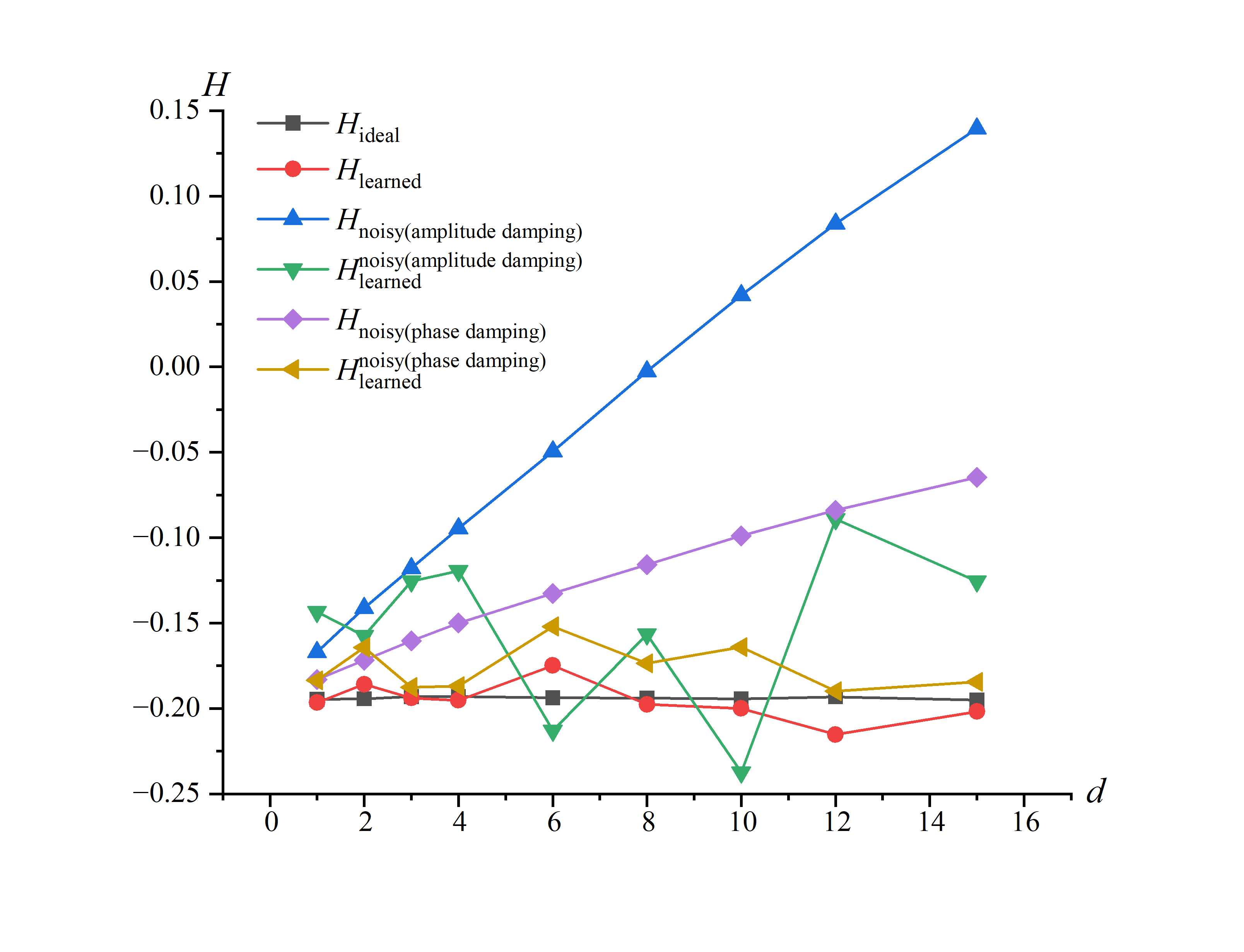}
    \caption{Hamiltonian $H$ at $x=0.3$ subjected to amplitude damping ($p=1\%$) and phase damping ($p=1\%$). The QCL‑based noise mitigation effectively suppresses the degradation, keeping the predicted values close to the exact conserved quantity.}
    \label{fig:H_amp_pha_x0_3}
\end{figure}

By measuring conserved charges, the effect of noise mitigation can be clearly visualized. In the presence of noise, conserved charges gradually decay as the time evolution progresses. However, when the effective circuit obtained through QCL is used for simulation, the conserved quantities can be maintained at a level close to their true values. Figure~\ref{fig:H_bit_dep_x0_3} and Figure~\ref{fig:H_amp_pha_x0_3} show the comparison among the ideal simulation of the original circuit, the ideal simulation of the learned circuit, the noisy simulation of the original circuit, and the noisy simulation of the learned circuit under different noise models. Each data point represents an average over 3 million measurements.

\begin{figure}[!htb]
    \centering
    \includegraphics[width=0.48\textwidth]{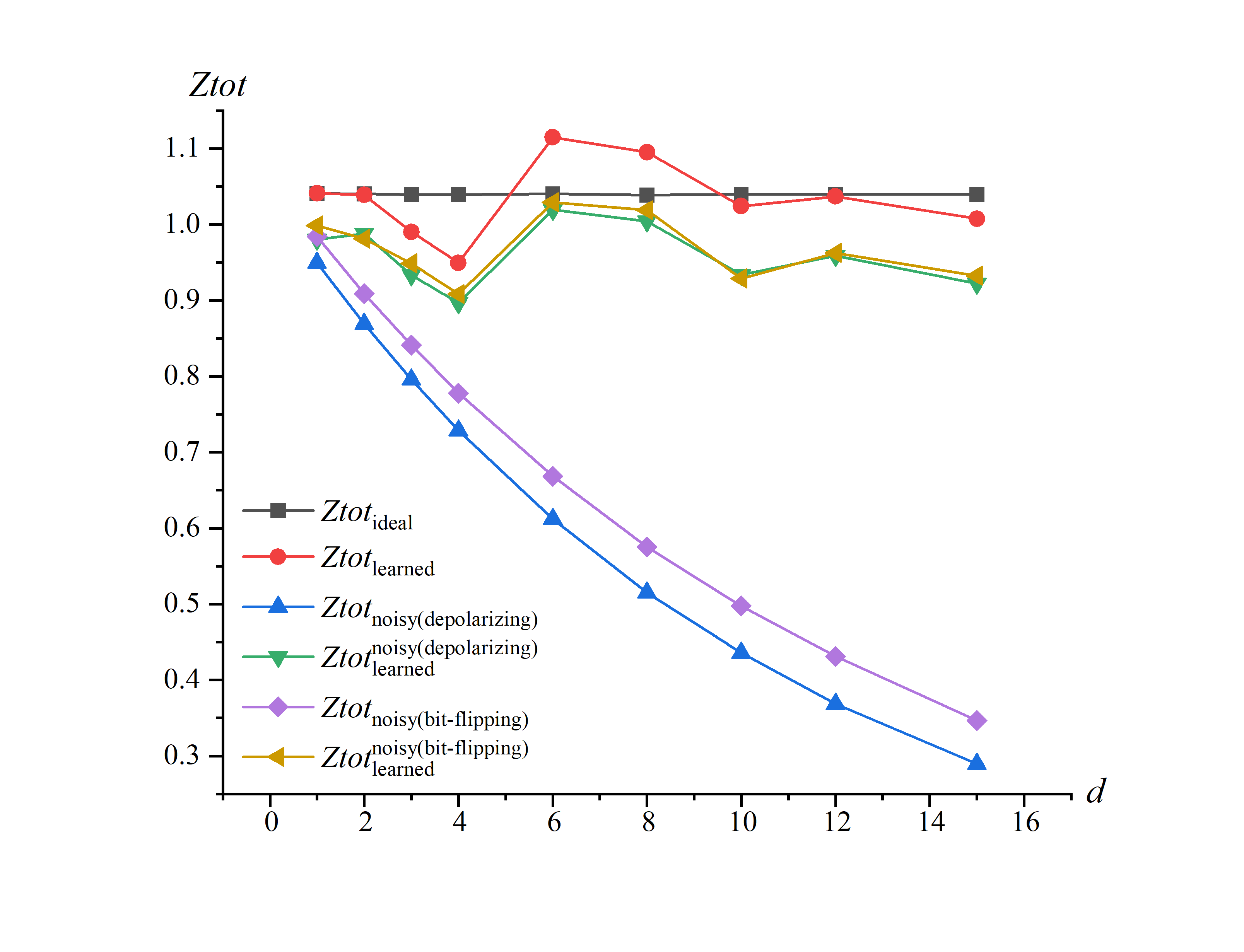}
    \caption{$Z_{\text{tot}}$ at $x=-0.3$ under bit‑flipping ($p=0.5\%$) and depolarizing ($p=1\%$) noise. Again, the learned circuit successfully limits the noise‑induced errors, maintaining the conserved charge much closer to its ideal value than the original time‑evolution circuit.}
    \label{fig:Ztot_bit_dep_xm0_3}
\end{figure}

\begin{figure}[!htb]
    \centering
    \includegraphics[width=0.48\textwidth]{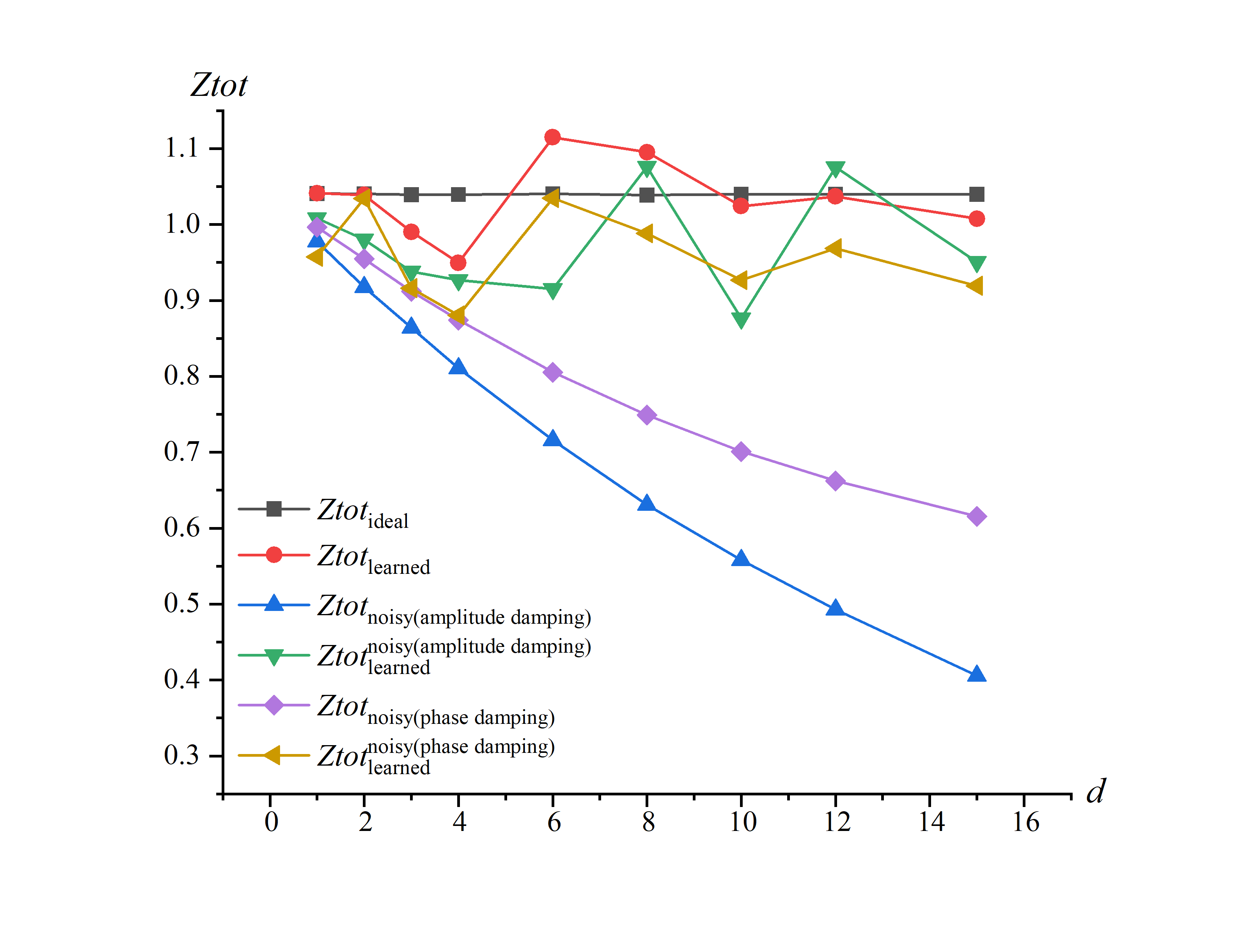}
    \caption{Total $Z$ magnetization $Z_{\text{tot}}$ at $x=-0.3$ under amplitude and phase damping noise. The learned circuit preserves the constant theoretical value with high accuracy, whereas the unmitigated circuit exhibits a large systematic drift caused by noise.}
    \label{fig:Ztot_amp_pha_xm0_3}
\end{figure}

Analogous benefits are observed for the total magnetization $Z_{\text{tot}}$. 
Figures~\ref{fig:Ztot_amp_pha_xm0_3} and~\ref{fig:Ztot_bit_dep_xm0_3} show that at a random input parameter $x=-0.3$, the learned circuit preserves $Z_{\text{tot}}$ well under all four noise types, essentially recovering the constant theoretical value across the full range of simulated depths up to $d=15$.

\subsection{Dynamic observables}

A primary goal of quantum simulation is to compute the time evolution of non-conserved observables. In this section, we demonstrate that the effective quantum circuit obtained via QCL can also mitigate the impact of noise in the simulation of such dynamical quantities. The learned circuit can reproduce key dynamical observables, such as $\langle \vec{\sigma}_1\rangle$, with significantly reduced noise-induced errors under realistic noise channels compared to the original simulation circuit.

As described before, only one of the dynamical observable, the expectation value $\langle Z_1\rangle$ of the first qubit, and the conserved charges are used as training data, but interestingly, the QCL circuit can predict the other dynamical observables rather precisely. Dynamical observables also undergo significant decay in noisy simulations, and this tendency shows little dependence on the specific type of noise. In the following, let us present some examples of the obtained results with $d=10$ (and $D=3$) fixed for different observables and different noisy models. 

\begin{figure}[!htb]
\centering
\includegraphics[width=0.48\textwidth]{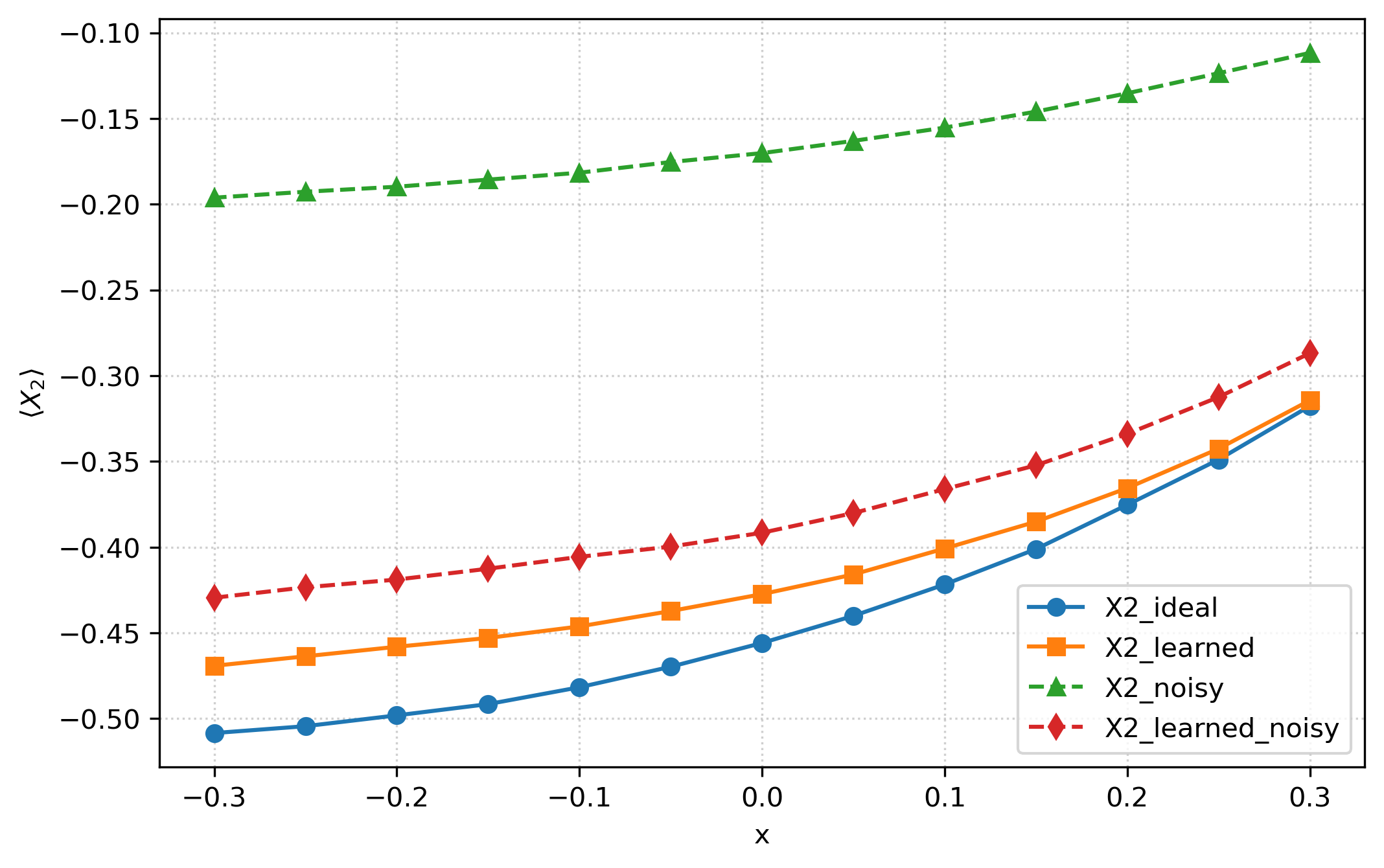}
\caption{Comparison of $\langle X_2\rangle$ as a function of $x$ after $d=10$ steps of time evolution between the ideal simulation (blue line for the original circuit, orange line for the prediction of QCL circuit) and the noisy simulation (green line for the original circuit, red line for the prediction of QCL circuit) with depolarizing noise ($p=0.01$).}
\label{fig:X1_depol}
\end{figure}

\begin{figure}[!htb]
\centering
\includegraphics[width=0.48\textwidth]{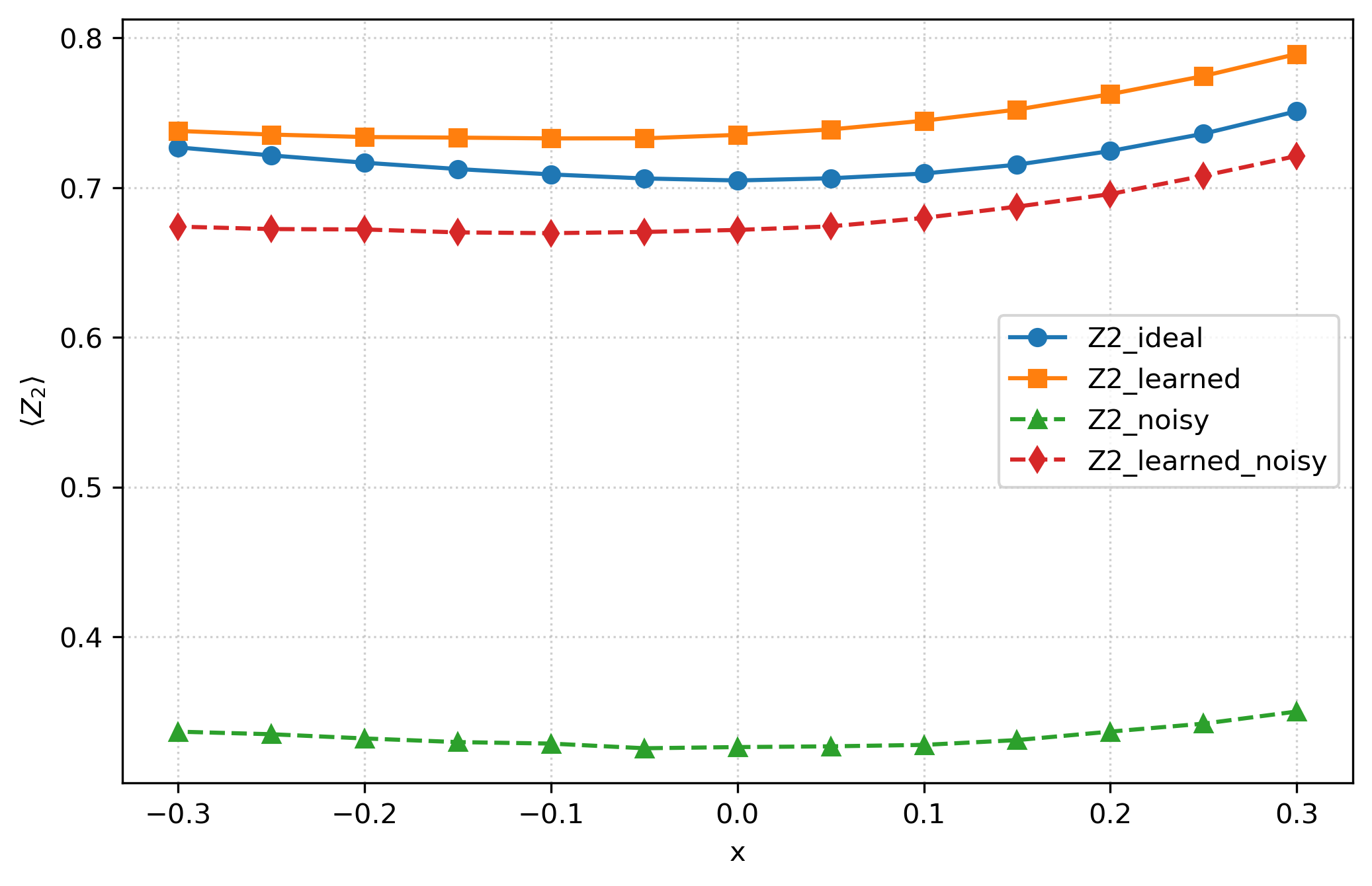}
\caption{Comparison of $\langle Z_2\rangle$ as a function of $x$ after $d=10$ steps of time evolution between the ideal simulation (blue line for the original circuit, orange line for the prediction of QCL circuit) and the noisy simulation (green line for the original circuit, red line for the prediction of QCL circuit) with bit-flipping noise ($p=0.005$).}
\label{fig:Z1_bitflip}
\end{figure}

Figure \ref{fig:X1_depol} shows the expectation value $\langle X_2\rangle$ under the ideal simulation and the simulation with depolarizing noise ($p=0.01$). In the noisy simulation of the original circuit, the value of $\langle X_2\rangle$ undergoes a substantial decay of over 50$\%$. In contrast, the circuit learned via QCL not only provides a closely matching prediction of $\langle X_2\rangle$, but also exhibits significantly reduced susceptibility to noise in the presence of the same noise channel. A similar observation is made in Figure \ref{fig:Z1_bitflip} for $\langle Z_2\rangle$ with bit-flipping noise ($p=0.005$), although the $z$-basis measurement is more sensitive to this type of noise. 

Amplitude damping noise models energy relaxation processes, which drive the system toward the ground state and predominantly degrade the off‑diagonal elements (coherence) of the density matrix. This type of noise is particularly detrimental to coherence and can severely distort observables that involve $y$‑basis measurements.

\begin{figure}[!htb]
\centering
\includegraphics[width=0.48\textwidth]{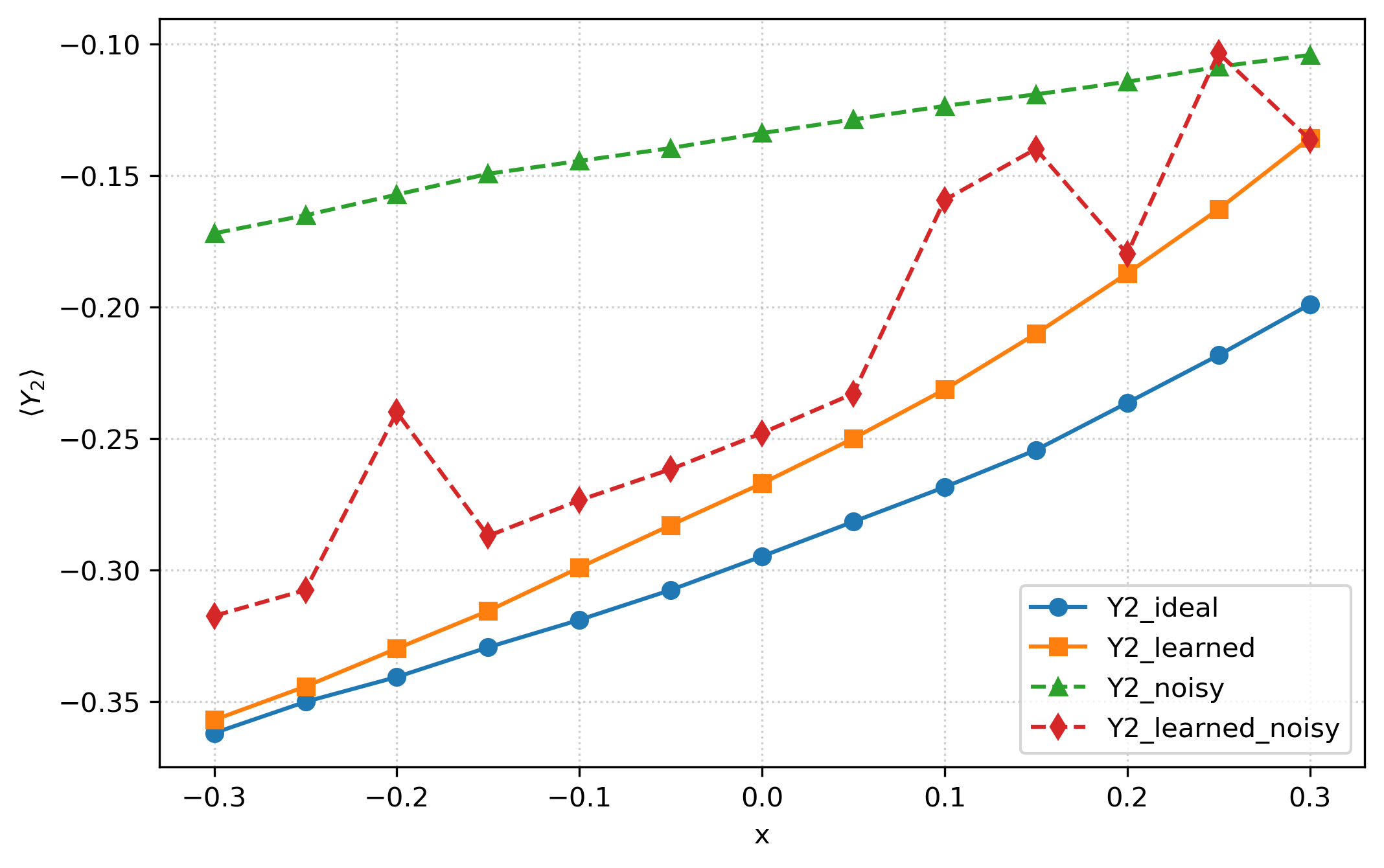}
\caption{Comparison of $\langle Y_2\rangle$ as a function of $x$ after $d=10$ steps of time evolution between the ideal simulation (blue line for the original circuit, orange line for the prediction of QCL circuit) and the noisy simulation (green line for the original circuit, red line for the prediction of QCL circuit) with amplitude damping noise ($p=0.01$).}
\label{fig:Y1_ampdamp}
\end{figure}

The effectiveness of the QCL-based noise mitigation under this noise channel is particularly notable, as this noise channel is non‑unital and induces an asymmetric contraction of the Bloch sphere. Figure \ref{fig:Y1_ampdamp} shows the comparison of the $\langle Y_2\rangle$ measurements among the ideal simulation, the prediction from the QCL circuit, and noisy simulations with amplitude damping noise ($p=0.01$). The QCL circuit yields a relatively less accurate prediction for $\langle Y_2\rangle$, however, its noisy simulation still shows a clear improvement over the original circuit, and we observed fluctuations in the noisy simulation results of the QCL circuit, whose cause remains unclear. This suggests that the QCL circuit might be particularly sensitive to the amplitude damping noise. 

\begin{figure}[!htb]
\centering
\includegraphics[width=0.48\textwidth]{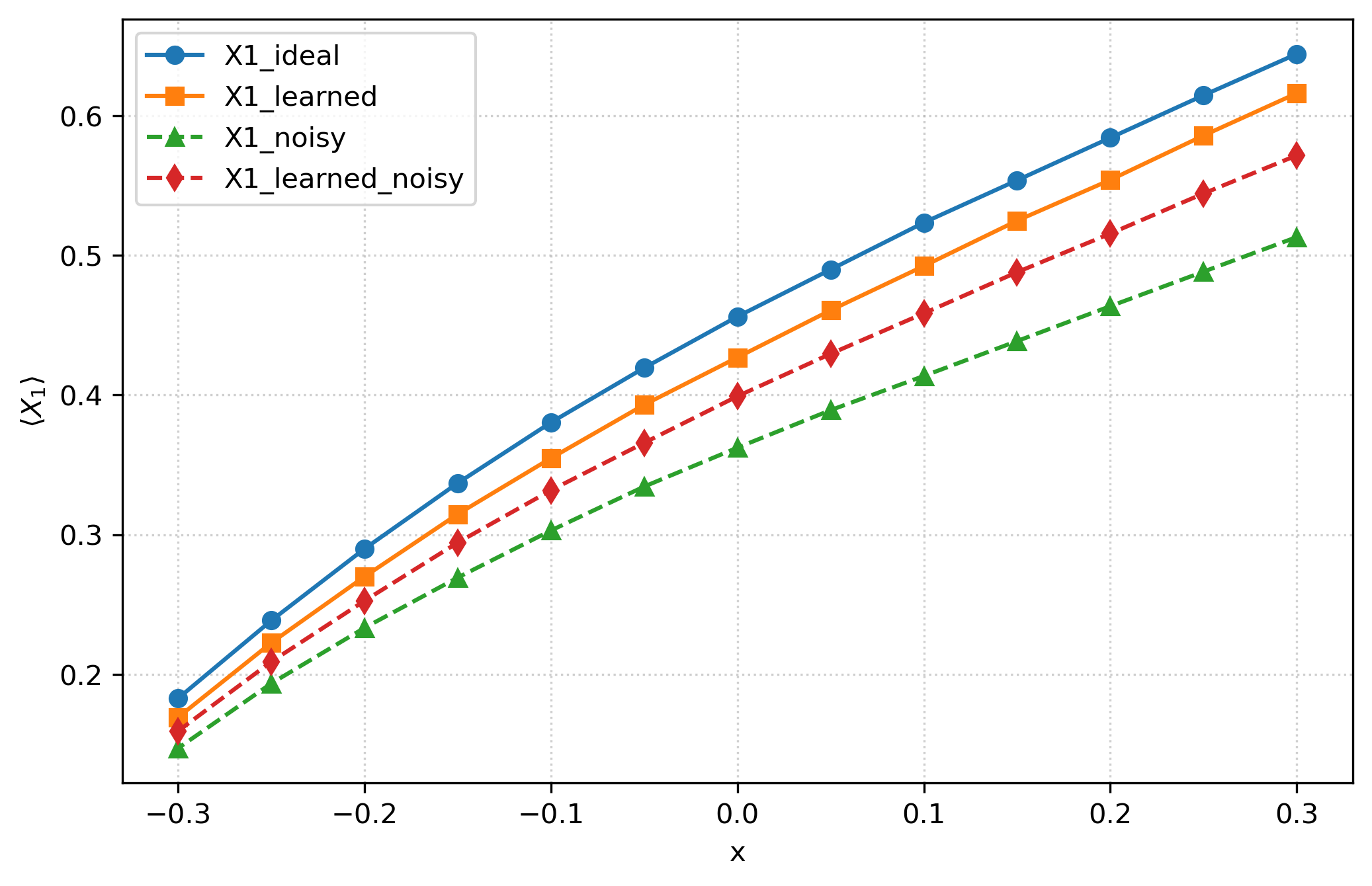}
\caption{Comparison of $\langle X_1\rangle$ as a function of $x$ after $d=10$ steps of time evolution between the ideal simulation (blue line for the original circuit, orange line for the prediction of QCL circuit) and the noisy simulation (green line for the original circuit, red line for the prediction of QCL circuit) with phase damping noise ($p=0.01$).}
\label{fig:X0_phasedamp}
\end{figure}

Phase damping noise represents pure decoherence without energy relaxation, specifically destroying the off-diagonal elements (coherence) of the density matrix in the computational basis while leaving the diagonal populations unchanged. This noise channel is particularly challenging for quantum simulation, as it directly degrades the quantum phase information essential for interference phenomena, with no straightforward classical analogue. Figure \ref{fig:X0_phasedamp} displays the simulation results for the measurements of $\langle X_1\rangle$ under phase damping noise ($p=0.01$). Even for this coherence-sensitive observable, the circuit learned via QCL demonstrates a measurable noise-reduction capability, producing predictions that are closer to the ideal noiseless values compared to the original simulation circuit under the same noise condition.

\subsection{Expressivity-noise trade-off and circuit depth}

In the noise mitigation approach presented in this work, a trade-off must be made between the expressivity of the circuit and its noise resilience. While deeper learning circuits generally achieve better training performance, they also become more vulnerable to the accumulation of noise. Our method addresses this by identifying a suitable circuit depth that keeps the noise-induced deviations at a stable and acceptable level. Notably, even relatively deep circuits with $D=4$ ($d\geq 6$) exhibit robust noise suppression, as reflected in the well-preserved behavior of conserved quantities. These results suggest that an optimal balance between expressivity and noise resilience can be achieved through a hyperparameter tuning strategy that warrants further systematic investigation in the future.

\subsection{Results for $L=3$ System}

To show the potential scalability of our QCL-based noise‑mitigation approach, we apply the same methodology to the $L=3$ Heisenberg spin chain. The system is evolved using the approximate time‑evolution operator \eqref{U3}, $U_3(\delta)$, and we use the conserved charge, the XXX Hamiltonian $H$, as a benchmark to assess the noise‑mitigation effect. The results shown here are based on the QCL circuit trained with a learning depth of $D=2$ for $d=1,2,3,5$, and noisy simulations are performed with depolarizing noise ($p=1\%$) ,bit‑flipping noise ($p=0.5\%$),amplitude
damping (\(p=1\%\)) and phase damping (\(p=1\%\)).

\begin{figure}[htb]
    \centering
    \includegraphics[width=0.48\textwidth]{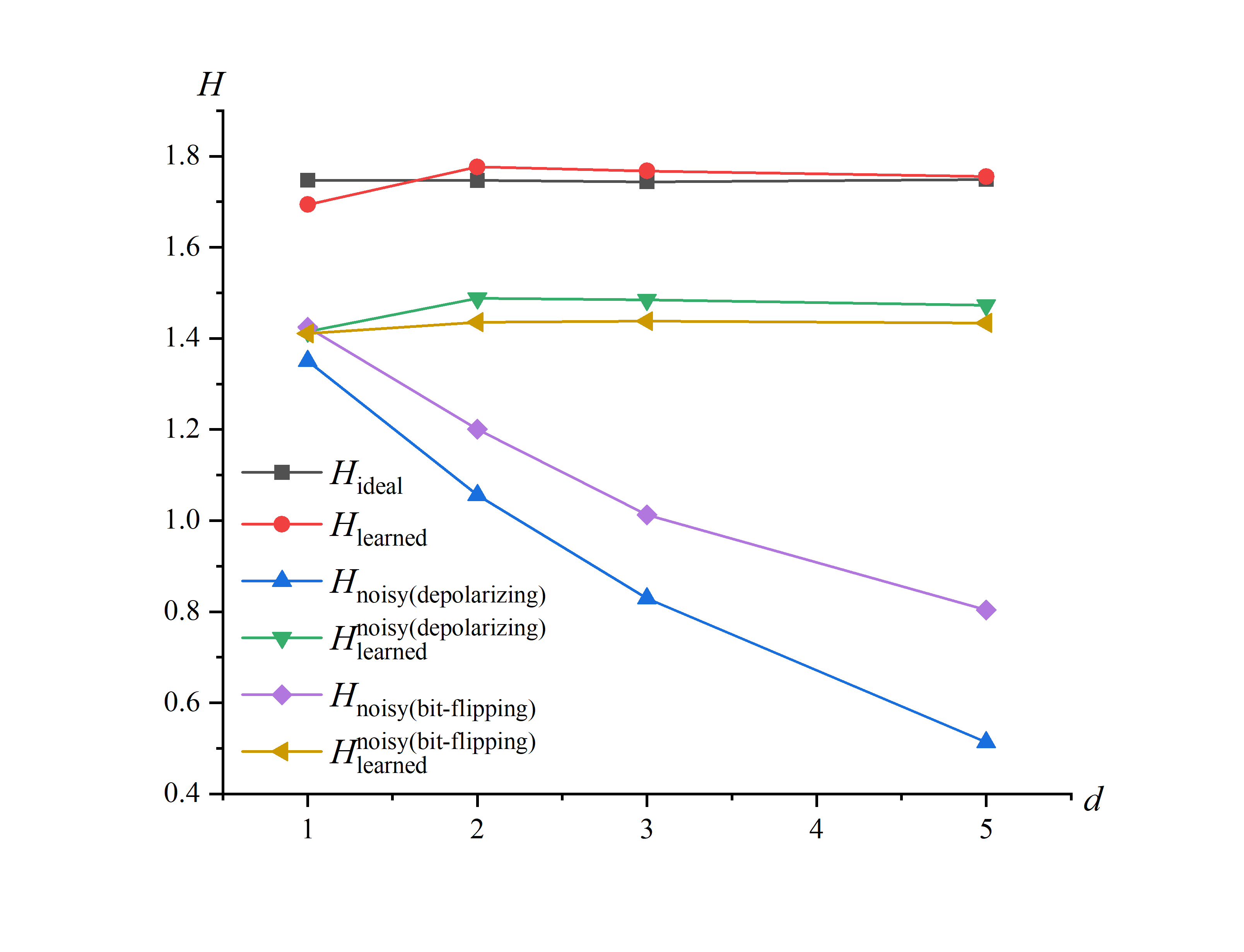}
    \caption{Hamiltonian \(H\) at \(x=-0.3\) for the three‑qubit system under bit‑flipping
    (\(p=0.5\%\)) and depolarizing (\(p=1\%\)) noise.  The learned circuit maintains the
    ideal value within \(\sim 15\%\) even at \(d=5\).}
    \label{fig:3qubit_H_bit_dep}
\end{figure}

\begin{figure}[htb]
    \centering
    \includegraphics[width=0.48\textwidth]{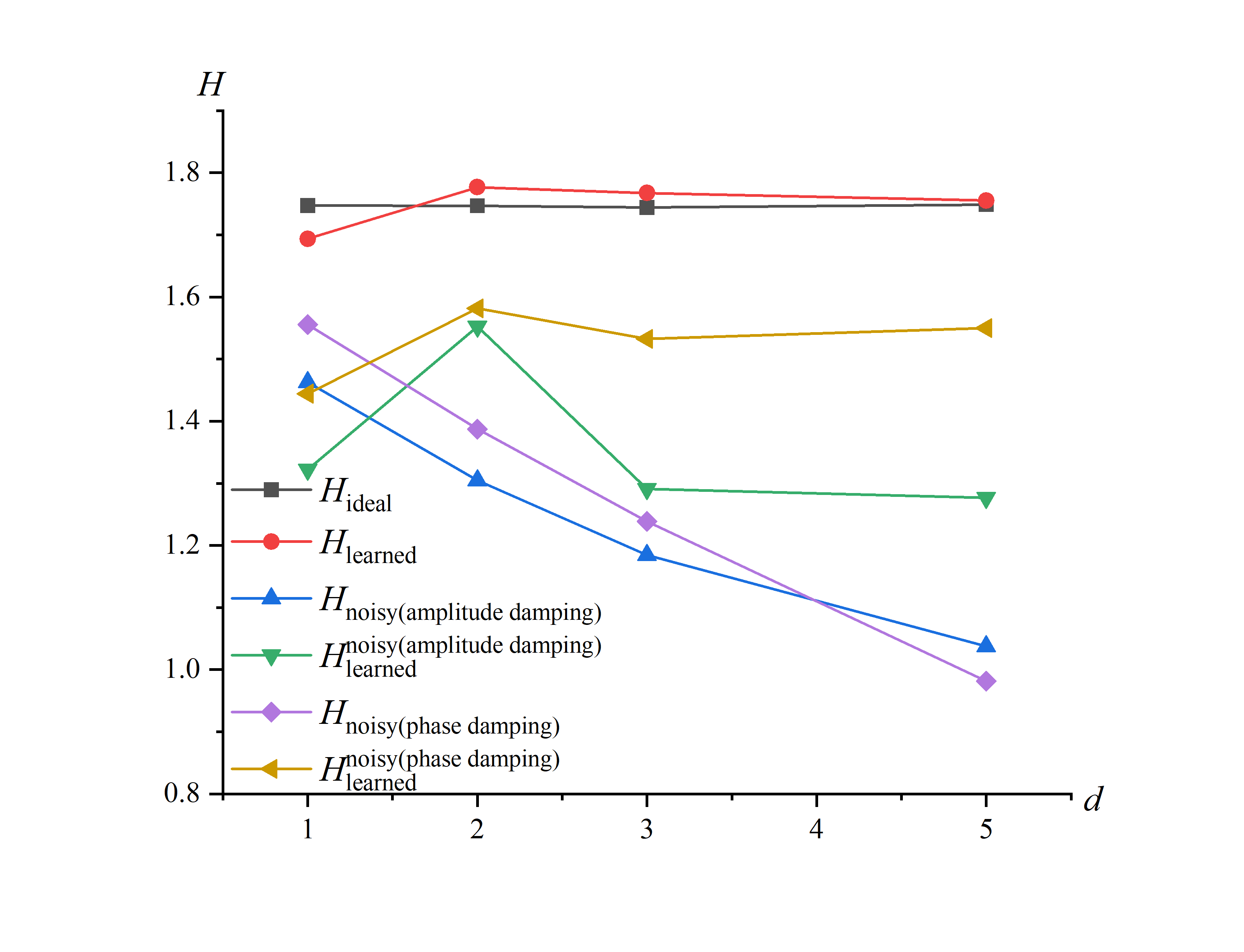}
    \caption{\(H\) at \(x=-0.3\) under amplitude damping (\(p=1\%\)) and phase damping
    (\(p=1\%\)).  The QCL‑trained circuit strongly suppresses the noise‑induced decay
    observed in the original circuit.}
    \label{fig:3qubit_H_amp_pha}
\end{figure}

Figure~\ref{fig:3qubit_H_bit_dep} compares the expectation value of the conserved Hamiltonian \(H\) at \(x=-0.3\) under bit‑flip (\(p=0.5\%\)) and depolarizing (\(p=1\%\)) noise.  The learned circuit remains very close to the ideal constant \(H_{\text{ideal}}\approx 1.747\), whereas the unmitigated circuit decays rapidly: already at \(d=5\) the noisy value drops below \(0.9\), corresponding to an error exceeding \(48\%\). Once again the QCL‑trained circuit effectively preserves the conserved charge, limiting the relative deviation to less than \(15\%\) at \(d=5\), while the unmitigated noisy simulation drifts substantially. The companion plot, Fig.~\ref{fig:3qubit_H_amp_pha}, shows the same observable under amplitude damping (\(p=1\%\)) and phase damping (\(p=1\%\)). Although the performance of the QCL circuit also degrades under these coherence-sensitive noise channels, it nevertheless maintains a clear advantage over the simulation using the original circuit, exhibiting significantly smaller deviations from the ideal noiseless values.

\begin{figure}[htb]
    \centering
    \includegraphics[width=0.48\textwidth]{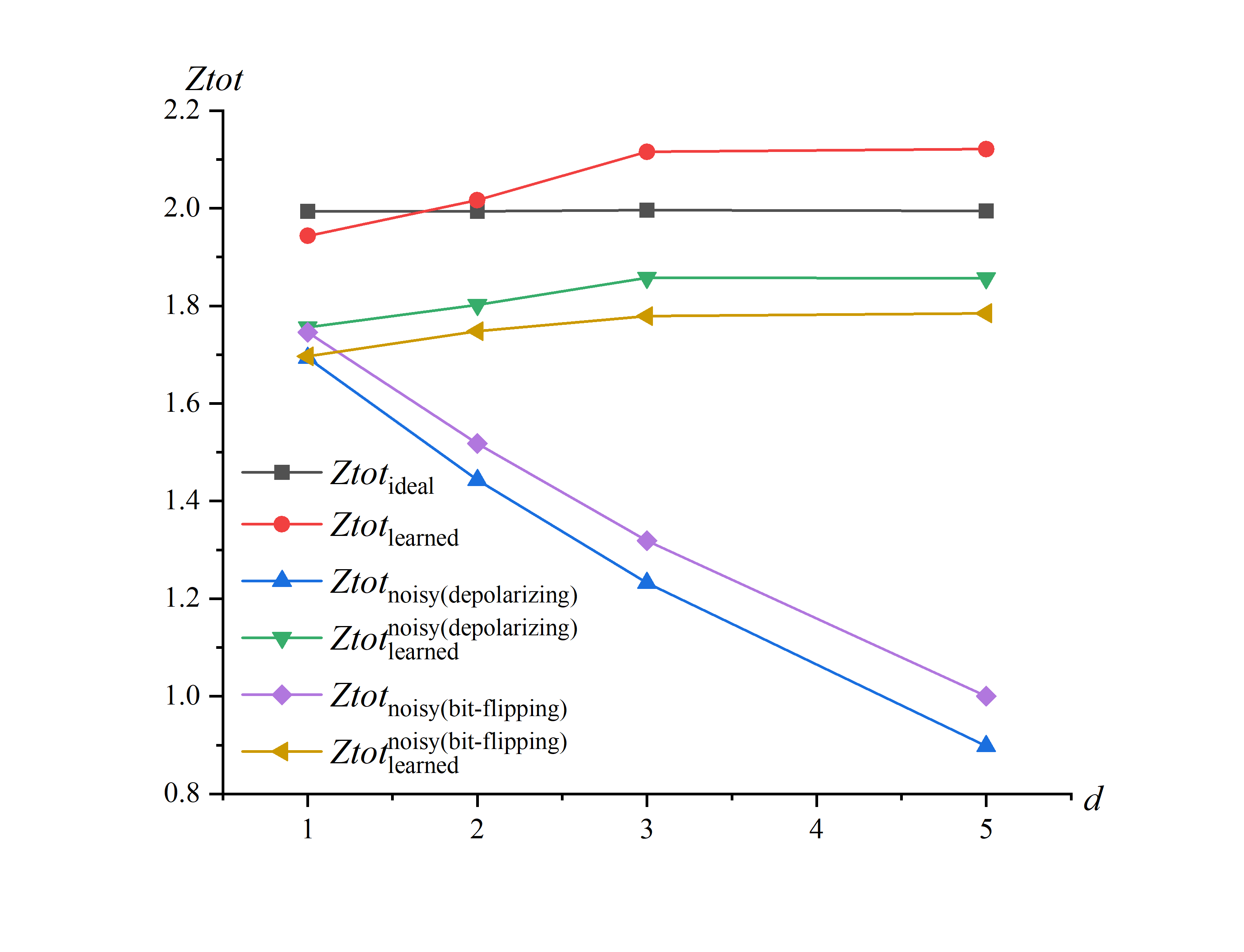}
    \caption{Total magnetization \(Z_{\text{tot}}\) at \(x=-0.3\) under bit‑flipping
    (\(p=0.5\%\)) and depolarizing (\(p=1\%\)) noise.  The learned circuit essentially
    recovers the ideal conserved value at all depths.}
    \label{fig:3qubit_Ztot_bit_dep}
\end{figure}

\begin{figure}[htb]
    \centering
    \includegraphics[width=0.48\textwidth]{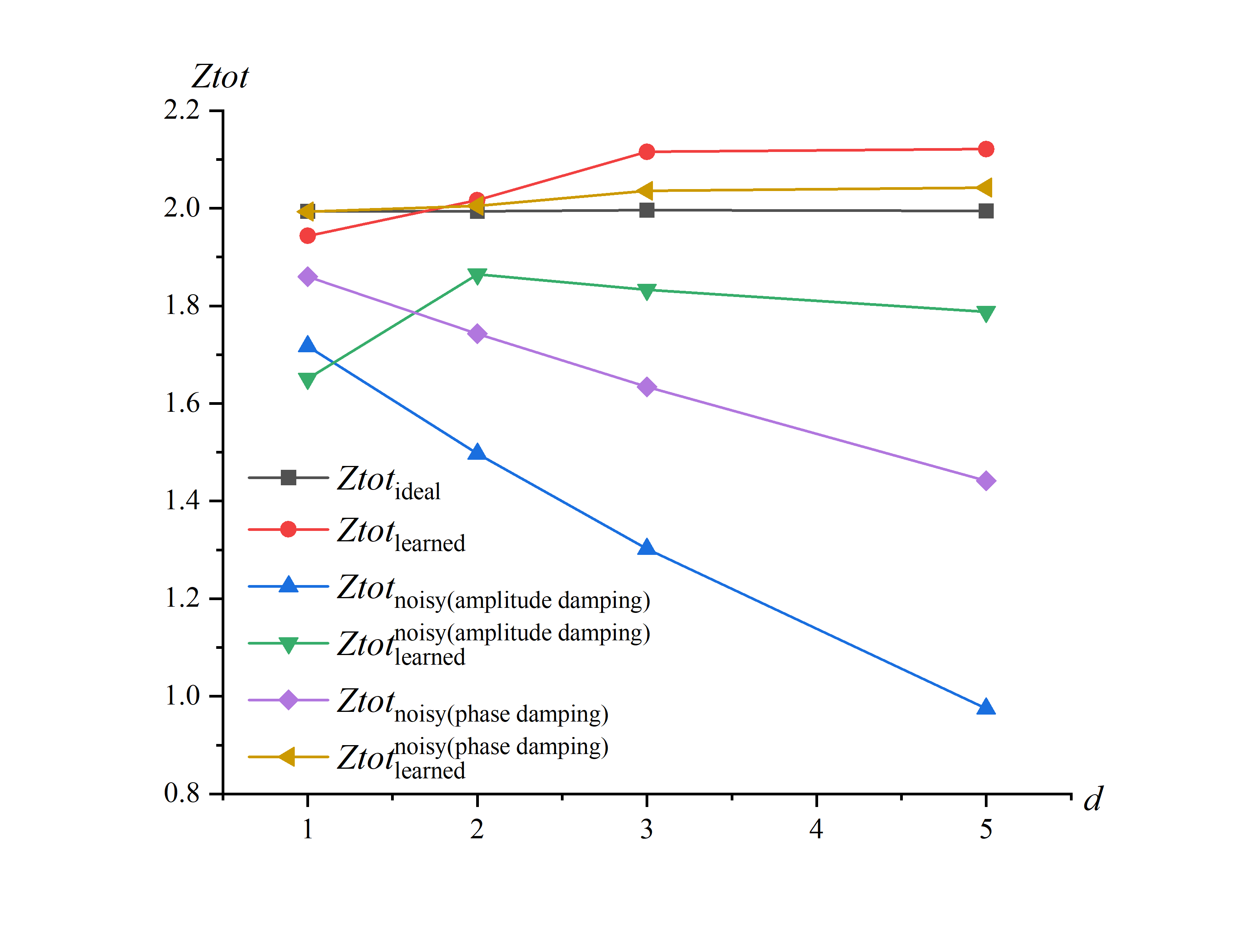}
    \caption{\(Z_{\text{tot}}\) at \(x=-0.3\) under amplitude damping (\(p=1\%\)) and
    phase damping (\(p=1\%\)).  The noise‑mitigation gain is independent of the error
    model.}
    \label{fig:3qubit_Ztot_amp_pha}
\end{figure}

\begin{figure}[htb]
    \centering
    \includegraphics[width=0.48\textwidth]{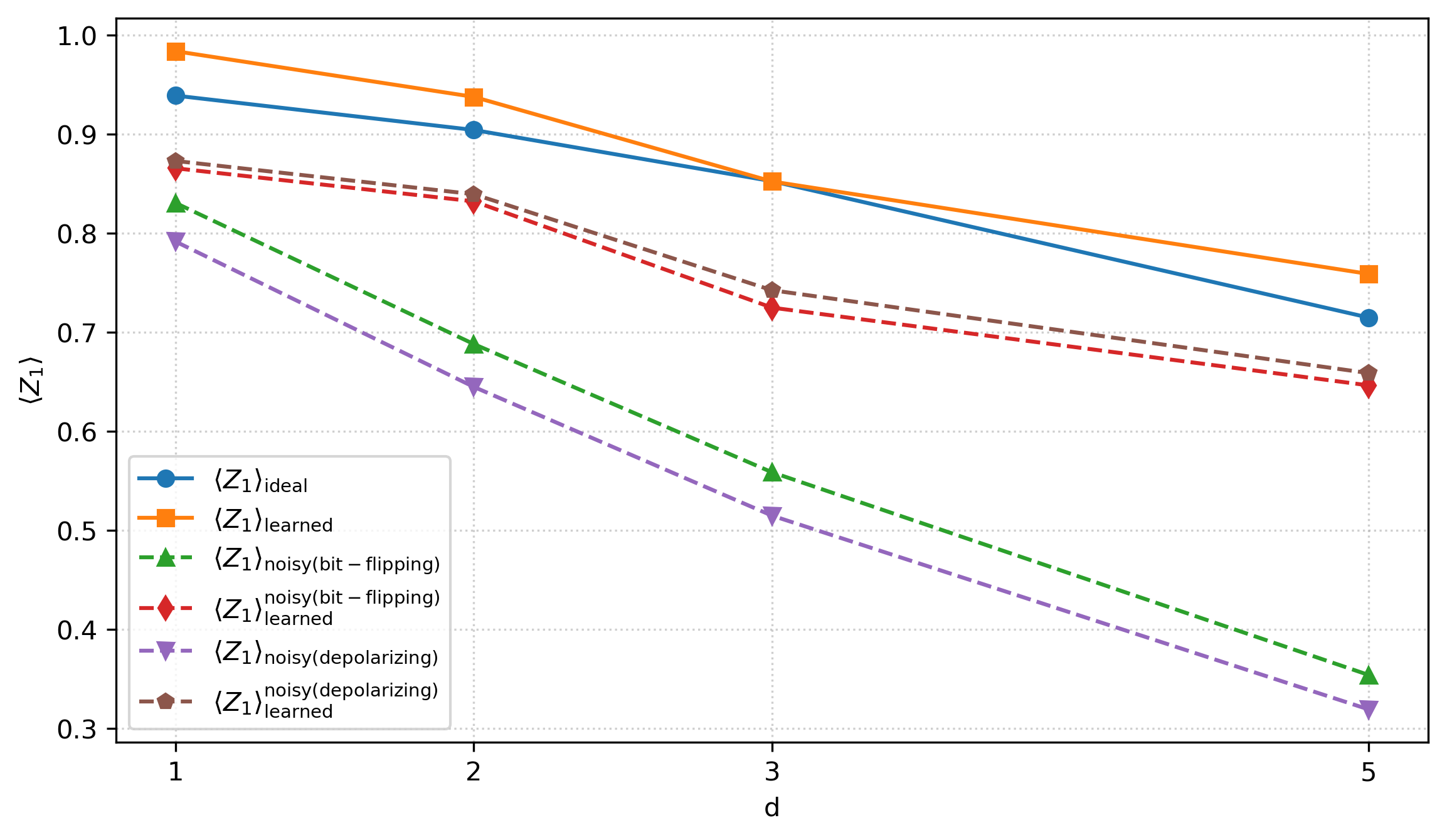}
    \caption{$\langle Z_1 \rangle$ at $x=-0.3$ under bit‑flipping and depolarizing
    noise.}
    \label{fig:3qubit_Z1_bit_dep}
\end{figure}

\begin{figure}[htb]
    \centering
    \includegraphics[width=0.48\textwidth]{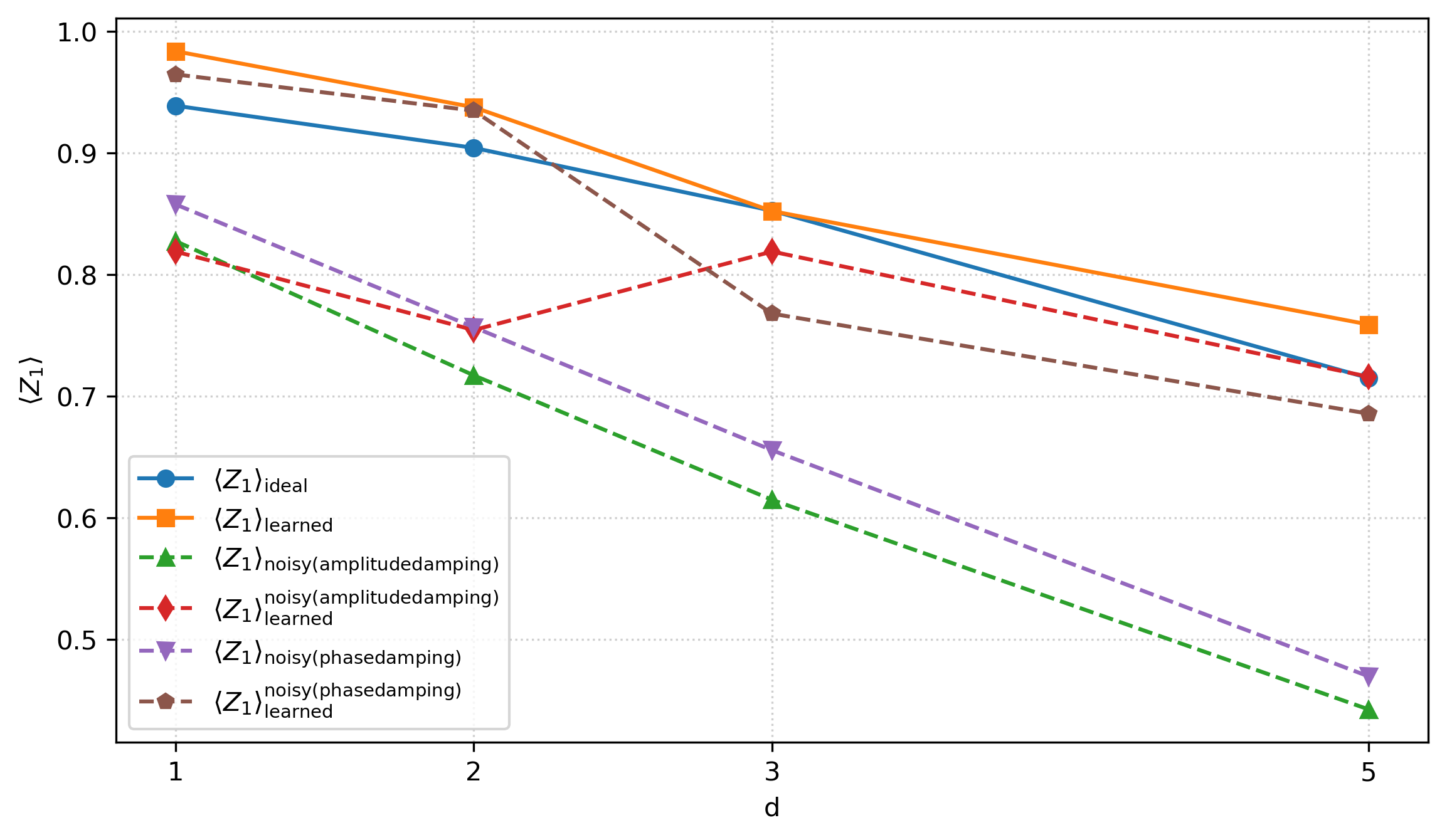}
    \caption{$\langle Z_1 \rangle$ at $x=-0.3$ under amplitude and phase damping
    noise.}
    \label{fig:3qubit_Z1_amp_pha}
\end{figure}

The QCL circuit exhibits the same level of protection for the total magnetization \(Z_{\text{tot}}=\sum_{i}Z_i\) and the dynamical observable $\langle Z_1\rangle$.  Figure~\ref{fig:3qubit_Ztot_bit_dep} and \ref{fig:3qubit_Z1_bit_dep} respectively display
\(Z_{\text{tot}}\) and $\langle Z_1\rangle$ in the presence of bit‑flip and depolarizing errors, and Fig.~\ref{fig:3qubit_Ztot_amp_pha} and \ref{fig:3qubit_Z1_amp_pha} cover the amplitude‑ and phase‑damping channels. In all cases the learned circuit reproduces the constant theoretical value (\(Z_{\text{tot}}\approx 1.994\)) and vary values of $\langle Z_1\rangle$ with good precision

These results demonstrate that the QCL-based noise‑mitigation strategy generalizes effectively beyond the two‑qubit prototype shown in details before. The learned circuits preserve the theoretically constant conserved charges to a much higher degree than the original circuits under the same noise levels, indicating that the approach remains effective when moving from two to three qubits and suggesting that it may extend to larger quantum systems.

\section{Conclusion and Discussion}\label{s:con}

Beyond the specific numerical examples presented in this work, our approach illustrates a general paradigm for physics-informed error mitigation in quantum simulation. By using conserved charges as both training data (or constraints) and diagnostic benchmarks, we obtain a shallow variational circuit that effectively compresses the deep time-evolution operator on a randomly-chosen subspace of physical states. The prototype results for $L = 2$ and $L = 3$ indicate that this strategy can significantly reduce noise-induced errors even under realistic decoherence channels, and suggest that incorporating more general symmetry structures and conserved quantities may further extend its applicability to larger and more complex quantum many-body systems.

Compared to conventional error-mitigation techniques like Zero-Noise Extrapolation (ZNE) and Probabilistic Error Cancellation (PEC) \cite{Temme:2016vkz} (also refer to \cite{Endo:2020kro,Koczor:2020wrs} and the discussion on theoretical limits \cite{Takagi:2021nnm}), the QCL-based strategy offers distinct practical advantages. It operates primarily as an offline, pre-computation step: once a noise-resilient circuit is learned from classical simulations, it can be executed on hardware a single time per inference point. This contrasts with ZNE, which requires multiple circuit runs at boosted noise levels, and PEC, which incurs a substantial sampling overhead. Therefore, QCL is particularly advantageous for simulating models with high symmetry, especially integrable models, where training data can be efficiently generated.

There are several future directions to be explored: 

\paragraph{$L=4$ and beyond} Extending the method to longer spin chains with $L=4 $ and beyond is conceptually straightforward. The primary technical consideration involves the measurement of higher-order conserved charges, whose explicit forms become increasingly complex. A more convenient alternative for training in these larger systems is to directly utilize the transfer matrix $T(u)$ at different spectral parameters $u_i$. One has to be careful that $T(u)$ is not a Hermitian operator, therefore, it must be decomposed into its Hermitian and anti-Hermitian parts (or equivalently, its real and imaginary components) and expressed in terms of Pauli matrices for separate training within the QCL framework. Based on this approach, we have conducted preliminary studies. The results indicate that the learned circuit can provide reasonable qualitative predictions for key dynamical observables such as $\langle X_1\rangle$ and $\langle Y_1\rangle$. It remains, however, as an open question on how to train such a learning circuit more efficiently and accurately. 

\paragraph{As variational quantum algorithm} Historically, the QCL was inspired by the foundational framework of the Variational Quantum Eigensolver (VQE) \cite{Peruzzo:2013bzg,Kandala:2017vok}, and in turn, the method presented in this work can also be utilized as a variational quantum algorithm. In this context, the method presented in this paper serves as a semi-classical (hybrid) strategy to solve for dynamical observables. In particular, it trains a compact circuit to encode the conserved charges and a sparse set of dynamical observables, which can then be used to predict more complex dynamical evolution. Related variational quantum simulation approaches have been developed in \cite{Yuan:2018jdl,Endo:2020sxk} to directly optimize a parameterized ansatz to match the time-dependent Schrödinger equation, whereas our method learns an effective evolution circuit by fitting conserved charges and a small set of dynamical observables, and then reuses it as a compressed simulator.

While promising, the current implementation leaves room for optimization. The original circuit for simulating $d$ steps of discrete time evolution requires $5Ld$ single-qubit gates and $4Ld$ CNOT gates. Constructing the learning circuit using the Hamiltonian $H_d$ in \eqref{H-train} demands $L(L+5)D/2$ single-qubit gates and $L(L-1)D$ CNOT gates. The key advantage of our approach is that for a fixed system size $L$, the required learning depth $D$ grows very slowly with the evolution depth $d$. This enables the representation of the same time-evolution operator with a shallower circuit, thereby reducing the impact of noise. However, for large $L$, the QCL circuit only becomes advantageous for relatively large $d$.

An important direction is to explore alternative choices for $H_d$ that maintain good learning performance while ensuring the gate count scales only linearly with $L$ \footnote{We performed a preliminary test with the nearest-neighbor Ising model Hamiltonian as $H_d$, but the QCL training result was not as good. }, or to apply our method to spin chains with long-range interaction. The Haldane-Shastry model \cite{Haldane:1987gg,SriramShastry:1987wdh} provides a good candidate for future study.

\begin{figure}[!htb]
\centering
\includegraphics[width=0.48\textwidth]{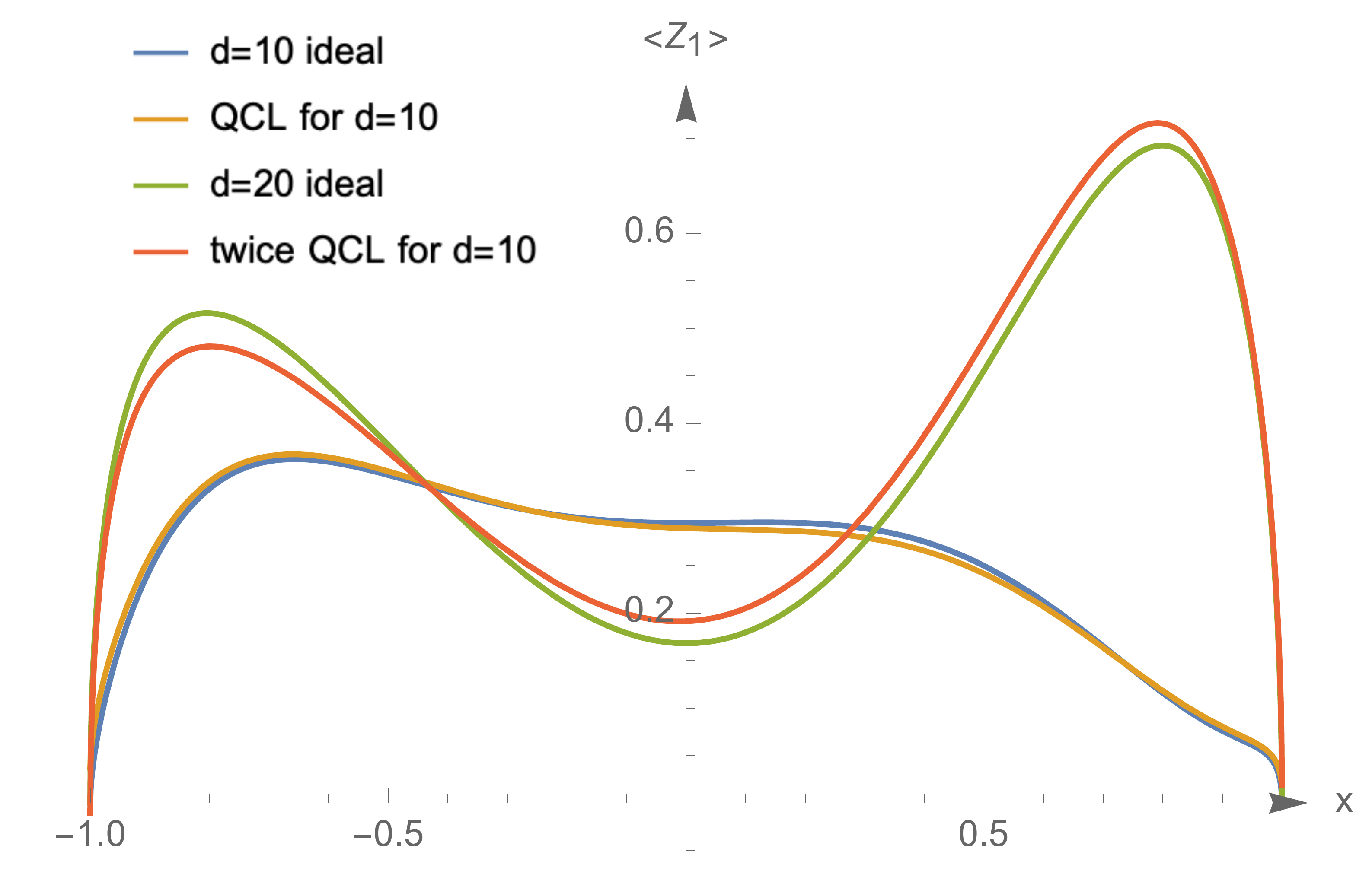}
\caption{Reusability of the learned circuit for extended time evolution. The expectation value of $\langle Z_1\rangle$ is shown for different simulation schemes. The curve labeled {\it twice QCL for $d=10$} is obtained by applying the circuit learned for 
$d=10$ evolution steps twice in sequence. Its close agreement with the theoretical curve labeled by {\it $d=20$ ideal} demonstrates that the noise-resilient, learned circuit can be reused to accurately simulate dynamics for twice the original time depth, validating a key advantage of our approach. }
\label{fig:pred-20}
\end{figure}

\paragraph{Towards a true quantum simulation} The ultimate goal of this method is, of course, its application to real quantum simulation. When the scaling factor $a$ is very close to $1$, the learned circuit achieves high fidelity and can be reused iteratively—meaning a circuit of depth $n\times D$ can simulate $n\times d$ time evolution steps. This would significantly shorten the overall simulation circuit and enable the prediction of long-time dynamics using observables trained on shorter evolution sequences. For the $L=2$ case, we have conducted preliminary verification of this concept (see Figure \ref{fig:pred-20}). However, how to effectively extend this approach to longer spin chains remains a subject for future investigation.

\subsection*{Acknowledgment} We would like to thank J. M. Kosterlitz for inspiring discussions. This work is partially supported by the National Natural Science Foundation of China (Grant No. 12105198).

\bibliography{QC}

\begin{thebibliography}{26}%
\makeatletter
\providecommand \@ifxundefined [1]{%
 \@ifx{#1\undefined}
}%
\providecommand \@ifnum [1]{%
 \ifnum #1\expandafter \@firstoftwo
 \else \expandafter \@secondoftwo
 \fi
}%
\providecommand \@ifx [1]{%
 \ifx #1\expandafter \@firstoftwo
 \else \expandafter \@secondoftwo
 \fi
}%
\providecommand \natexlab [1]{#1}%
\providecommand \enquote  [1]{``#1''}%
\providecommand \bibnamefont  [1]{#1}%
\providecommand \bibfnamefont [1]{#1}%
\providecommand \citenamefont [1]{#1}%
\providecommand \href@noop [0]{\@secondoftwo}%
\providecommand \href [0]{\begingroup \@sanitize@url \@href}%
\providecommand \@href[1]{\@@startlink{#1}\@@href}%
\providecommand \@@href[1]{\endgroup#1\@@endlink}%
\providecommand \@sanitize@url [0]{\catcode `\\12\catcode `\$12\catcode `\&12\catcode `\#12\catcode `\^12\catcode `\_12\catcode `\%12\relax}%
\providecommand \@@startlink[1]{}%
\providecommand \@@endlink[0]{}%
\providecommand \url  [0]{\begingroup\@sanitize@url \@url }%
\providecommand \@url [1]{\endgroup\@href {#1}{\urlprefix }}%
\providecommand \urlprefix  [0]{URL }%
\providecommand \Eprint [0]{\href }%
\providecommand \doibase [0]{http://dx.doi.org/}%
\providecommand \selectlanguage [0]{\@gobble}%
\providecommand \bibinfo  [0]{\@secondoftwo}%
\providecommand \bibfield  [0]{\@secondoftwo}%
\providecommand \translation [1]{[#1]}%
\providecommand \BibitemOpen [0]{}%
\providecommand \bibitemStop [0]{}%
\providecommand \bibitemNoStop [0]{.\EOS\space}%
\providecommand \EOS [0]{\spacefactor3000\relax}%
\providecommand \BibitemShut  [1]{\csname bibitem#1\endcsname}%
\let\auto@bib@innerbib\@empty
\bibitem [{\citenamefont {Mitarai}\ \emph {et~al.}(2018)\citenamefont {Mitarai}, \citenamefont {Negoro}, \citenamefont {Kitagawa},\ and\ \citenamefont {Fujii}}]{Mitarai:2018voy}%
  \BibitemOpen
  \bibfield  {author} {\bibinfo {author} {\bibfnamefont {K.}~\bibnamefont {Mitarai}}, \bibinfo {author} {\bibfnamefont {M.}~\bibnamefont {Negoro}}, \bibinfo {author} {\bibfnamefont {M.}~\bibnamefont {Kitagawa}}, \ and\ \bibinfo {author} {\bibfnamefont {K.}~\bibnamefont {Fujii}},\ }\href {\doibase 10.1103/PhysRevA.98.032309} {\bibfield  {journal} {\bibinfo  {journal} {Phys. Rev. A}\ }\textbf {\bibinfo {volume} {98}},\ \bibinfo {pages} {032309} (\bibinfo {year} {2018})},\ \Eprint {http://arxiv.org/abs/1803.00745}{arXiv:1803.00745 [quant-ph]}\BibitemShut {NoStop}%
\bibitem [{\citenamefont {Havlíček}\ \emph {et~al.}(2019)\citenamefont {Havlíček}, \citenamefont {Córcoles}, \citenamefont {Temme}, \citenamefont {Harrow}, \citenamefont {Kandala}, \citenamefont {Chow},\ and\ \citenamefont {Gambetta}}]{Havlicek2019}%
  \BibitemOpen
  \bibfield  {author} {\bibinfo {author} {\bibfnamefont {V.}~\bibnamefont {Havlíček}}, \bibinfo {author} {\bibfnamefont {A.~D.}\ \bibnamefont {Córcoles}}, \bibinfo {author} {\bibfnamefont {K.}~\bibnamefont {Temme}}, \bibinfo {author} {\bibfnamefont {A.~W.}\ \bibnamefont {Harrow}}, \bibinfo {author} {\bibfnamefont {A.}~\bibnamefont {Kandala}}, \bibinfo {author} {\bibfnamefont {J.~M.}\ \bibnamefont {Chow}}, \ and\ \bibinfo {author} {\bibfnamefont {J.~M.}\ \bibnamefont {Gambetta}},\ }\href {\doibase 10.1038/s41586-019-0980-2} {\bibfield  {journal} {\bibinfo  {journal} {Nature}\ }\textbf {\bibinfo {volume} {567}},\ \bibinfo {pages} {209} (\bibinfo {year} {2019})}\BibitemShut {NoStop}%
\bibitem [{\citenamefont {Li}\ and\ \citenamefont {Deng}(2021)}]{Li2021}%
  \BibitemOpen
  \bibfield  {author} {\bibinfo {author} {\bibfnamefont {W.}~\bibnamefont {Li}}\ and\ \bibinfo {author} {\bibfnamefont {D.-L.}\ \bibnamefont {Deng}},\ }\href {\doibase 10.1007/s11433-021-1793-6} {\bibfield  {journal} {\bibinfo  {journal} {Science China Physics, Mechanics \& Astronomy}\ }\textbf {\bibinfo {volume} {65}},\ \bibinfo {pages} {220301} (\bibinfo {year} {2021})}\BibitemShut {NoStop}%
\bibitem [{\citenamefont {Schnabel}\ and\ \citenamefont {Roth}(2025)}]{Schnabel2025}%
  \BibitemOpen
  \bibfield  {author} {\bibinfo {author} {\bibfnamefont {J.}~\bibnamefont {Schnabel}}\ and\ \bibinfo {author} {\bibfnamefont {M.}~\bibnamefont {Roth}},\ }\href {\doibase 10.1007/s42484-025-00273-5} {\bibfield  {journal} {\bibinfo  {journal} {Quantum Machine Intelligence}\ }\textbf {\bibinfo {volume} {7}},\ \bibinfo {pages} {58} (\bibinfo {year} {2025})}\BibitemShut {NoStop}%
\bibitem [{\citenamefont {Suzuki}\ \emph {et~al.}(2024)\citenamefont {Suzuki}, \citenamefont {Hasebe},\ and\ \citenamefont {Miyazaki}}]{Suzuki2024}%
  \BibitemOpen
  \bibfield  {author} {\bibinfo {author} {\bibfnamefont {T.}~\bibnamefont {Suzuki}}, \bibinfo {author} {\bibfnamefont {T.}~\bibnamefont {Hasebe}}, \ and\ \bibinfo {author} {\bibfnamefont {T.}~\bibnamefont {Miyazaki}},\ }\href {\doibase 10.1007/s42484-024-00165-0} {\bibfield  {journal} {\bibinfo  {journal} {Quantum Machine Intelligence}\ }\textbf {\bibinfo {volume} {6}},\ \bibinfo {pages} {31} (\bibinfo {year} {2024})}\BibitemShut {NoStop}%
\bibitem [{\citenamefont {Heimann}\ \emph {et~al.}(2025)\citenamefont {Heimann}, \citenamefont {Sch{\"o}nhoff}, \citenamefont {Mounzer}, \citenamefont {Hohenfeld},\ and\ \citenamefont {Kirchner}}]{Heimann:2022ytr}%
  \BibitemOpen
  \bibfield  {author} {\bibinfo {author} {\bibfnamefont {D.}~\bibnamefont {Heimann}}, \bibinfo {author} {\bibfnamefont {G.}~\bibnamefont {Sch{\"o}nhoff}}, \bibinfo {author} {\bibfnamefont {E.}~\bibnamefont {Mounzer}}, \bibinfo {author} {\bibfnamefont {H.}~\bibnamefont {Hohenfeld}}, \ and\ \bibinfo {author} {\bibfnamefont {F.}~\bibnamefont {Kirchner}},\ }\href {\doibase 10.1103/PhysRevResearch.7.023151} {\bibfield  {journal} {\bibinfo  {journal} {Phys. Rev. Res.}\ }\textbf {\bibinfo {volume} {7}},\ \bibinfo {pages} {023151} (\bibinfo {year} {2025})},\ \Eprint {http://arxiv.org/abs/2209.10345}{arXiv:2209.10345 [quant-ph]}\BibitemShut {NoStop}%
\bibitem [{\citenamefont {Maruyoshi}\ \emph {et~al.}(2023)\citenamefont {Maruyoshi}, \citenamefont {Okuda}, \citenamefont {Pedersen}, \citenamefont {Suzuki}, \citenamefont {Yamazaki},\ and\ \citenamefont {Yoshida}}]{Maruyoshi:2022jnr}%
  \BibitemOpen
  \bibfield  {author} {\bibinfo {author} {\bibfnamefont {K.}~\bibnamefont {Maruyoshi}}, \bibinfo {author} {\bibfnamefont {T.}~\bibnamefont {Okuda}}, \bibinfo {author} {\bibfnamefont {J.~W.}\ \bibnamefont {Pedersen}}, \bibinfo {author} {\bibfnamefont {R.}~\bibnamefont {Suzuki}}, \bibinfo {author} {\bibfnamefont {M.}~\bibnamefont {Yamazaki}}, \ and\ \bibinfo {author} {\bibfnamefont {Y.}~\bibnamefont {Yoshida}},\ }\href {\doibase 10.1088/1751-8121/acc369} {\bibfield  {journal} {\bibinfo  {journal} {J. Phys. A}\ }\textbf {\bibinfo {volume} {56}},\ \bibinfo {pages} {165301} (\bibinfo {year} {2023})},\ \Eprint {http://arxiv.org/abs/2208.00576}{arXiv:2208.00576 [quant-ph]}\BibitemShut {NoStop}%
\bibitem [{\citenamefont {Lloyd}(1996)}]{Lloyd:1996aai}%
  \BibitemOpen
  \bibfield  {author} {\bibinfo {author} {\bibfnamefont {S.}~\bibnamefont {Lloyd}},\ }\href {\doibase 10.1126/science.273.5278.1073} {\bibfield  {journal} {\bibinfo  {journal} {Science}\ }\textbf {\bibinfo {volume} {273}},\ \bibinfo {pages} {1073} (\bibinfo {year} {1996})}\BibitemShut {NoStop}%
\bibitem [{\citenamefont {Bethe}(1931)}]{Bethe:1931hc}%
  \BibitemOpen
  \bibfield  {author} {\bibinfo {author} {\bibfnamefont {H.}~\bibnamefont {Bethe}},\ }\href {\doibase 10.1007/BF01341708} {\bibfield  {journal} {\bibinfo  {journal} {Z. Phys.}\ }\textbf {\bibinfo {volume} {71}},\ \bibinfo {pages} {205} (\bibinfo {year} {1931})}\BibitemShut {NoStop}%
\bibitem [{\citenamefont {Faddeev}(1995)}]{Faddeev:1994nk}%
  \BibitemOpen
  \bibfield  {author} {\bibinfo {author} {\bibfnamefont {L.~D.}\ \bibnamefont {Faddeev}},\ }\href {\doibase 10.1142/S0217751X95000905} {\bibfield  {journal} {\bibinfo  {journal} {Int. J. Mod. Phys. A}\ }\textbf {\bibinfo {volume} {10}},\ \bibinfo {pages} {1845} (\bibinfo {year} {1995})},\ \Eprint {http://arxiv.org/abs/hep-th/9404013}{arXiv:hep-th/9404013}\BibitemShut {NoStop}%
\bibitem [{\citenamefont {Slavnov}(2019)}]{Slavnov:2018kfx}%
  \BibitemOpen
  \bibfield  {author} {\bibinfo {author} {\bibfnamefont {N.~A.}\ \bibnamefont {Slavnov}},\ }\href {https://arxiv.org/abs/1804.07350} {\enquote {\bibinfo {title} {Algebraic bethe ansatz},}\ } (\bibinfo {year} {2019}),\ \Eprint {http://arxiv.org/abs/1804.07350}{arXiv:1804.07350 [math-ph]}\BibitemShut {NoStop}%
\bibitem [{\citenamefont {{Vanicat}}\ \emph {et~al.}(2018)\citenamefont {{Vanicat}}, \citenamefont {{Zadnik}},\ and\ \citenamefont {{Prosen}}}]{2018PhRvL.121c0606V}%
  \BibitemOpen
  \bibfield  {author} {\bibinfo {author} {\bibfnamefont {M.}~\bibnamefont {{Vanicat}}}, \bibinfo {author} {\bibfnamefont {L.}~\bibnamefont {{Zadnik}}}, \ and\ \bibinfo {author} {\bibfnamefont {T.}~\bibnamefont {{Prosen}}},\ }\href {\doibase 10.1103/PhysRevLett.121.030606} {\bibfield  {journal} {\bibinfo  {journal} {Phys. Rev. Lett.}\ }\textbf {\bibinfo {volume} {121}},\ \bibinfo {eid} {030606} (\bibinfo {year} {2018})},\ \Eprint {http://arxiv.org/abs/1712.00431}{arXiv:1712.00431 [cond-mat.stat-mech]}\BibitemShut {NoStop}%
\bibitem [{\citenamefont {Khatri}\ \emph {et~al.}(2019)\citenamefont {Khatri}, \citenamefont {LaRose}, \citenamefont {Poremba}, \citenamefont {Cincio}, \citenamefont {Sornborger},\ and\ \citenamefont {Coles}}]{Khatri:2018mjs}%
  \BibitemOpen
  \bibfield  {author} {\bibinfo {author} {\bibfnamefont {S.}~\bibnamefont {Khatri}}, \bibinfo {author} {\bibfnamefont {R.}~\bibnamefont {LaRose}}, \bibinfo {author} {\bibfnamefont {A.}~\bibnamefont {Poremba}}, \bibinfo {author} {\bibfnamefont {L.}~\bibnamefont {Cincio}}, \bibinfo {author} {\bibfnamefont {A.~T.}\ \bibnamefont {Sornborger}}, \ and\ \bibinfo {author} {\bibfnamefont {P.~J.}\ \bibnamefont {Coles}},\ }\href {\doibase 10.22331/q-2019-05-13-140} {\bibfield  {journal} {\bibinfo  {journal} {Quantum}\ }\textbf {\bibinfo {volume} {3}},\ \bibinfo {pages} {140} (\bibinfo {year} {2019})},\ \Eprint {http://arxiv.org/abs/1807.00800}{arXiv:1807.00800 [quant-ph]}\BibitemShut {NoStop}%
\bibitem [{\citenamefont {Coles}\ \emph {et~al.}(2018)\citenamefont {Coles}, \citenamefont {Sornborger}, \citenamefont {Suba{\c{s}}{\i}},\ and\ \citenamefont {Cincio}}]{Coles:2018vmk}%
  \BibitemOpen
  \bibfield  {author} {\bibinfo {author} {\bibfnamefont {P.~J.}\ \bibnamefont {Coles}}, \bibinfo {author} {\bibfnamefont {A.~T.}\ \bibnamefont {Sornborger}}, \bibinfo {author} {\bibfnamefont {Y.}~\bibnamefont {Suba{\c{s}}{\i}}}, \ and\ \bibinfo {author} {\bibfnamefont {L.}~\bibnamefont {Cincio}},\ }\href {\doibase 10.1088/1367-2630/aae94a} {\bibfield  {journal} {\bibinfo  {journal} {New J. Phys.}\ }\textbf {\bibinfo {volume} {20}},\ \bibinfo {pages} {113022} (\bibinfo {year} {2018})},\ \Eprint {http://arxiv.org/abs/1803.04114}{arXiv:1803.04114 [quant-ph]}\BibitemShut {NoStop}%
\bibitem [{Note1()}]{Note1}%
  \BibitemOpen
  \bibinfo {note} {For $d=15$, the learning circuit with $D=4$ gives better results than $D=3$, so $D=4$ circuit was used in the noisy simulation. The learning result for $d=15$ and $D=3$ appears in Figure \ref {fig:learn}, and we include it for comparison of the overlap.}\BibitemShut {Stop}%
\bibitem [{\citenamefont {Temme}\ \emph {et~al.}(2017)\citenamefont {Temme}, \citenamefont {Bravyi},\ and\ \citenamefont {Gambetta}}]{Temme:2016vkz}%
  \BibitemOpen
  \bibfield  {author} {\bibinfo {author} {\bibfnamefont {K.}~\bibnamefont {Temme}}, \bibinfo {author} {\bibfnamefont {S.}~\bibnamefont {Bravyi}}, \ and\ \bibinfo {author} {\bibfnamefont {J.~M.}\ \bibnamefont {Gambetta}},\ }\href {\doibase 10.1103/physrevlett.119.180509} {\bibfield  {journal} {\bibinfo  {journal} {Phys. Rev. Lett.}\ }\textbf {\bibinfo {volume} {119}},\ \bibinfo {pages} {180509} (\bibinfo {year} {2017})},\ \Eprint {http://arxiv.org/abs/1612.02058}{arXiv:1612.02058 [quant-ph]}\BibitemShut {NoStop}%
\bibitem [{\citenamefont {Endo}\ \emph {et~al.}(2021)\citenamefont {Endo}, \citenamefont {Cai}, \citenamefont {Benjamin},\ and\ \citenamefont {Yuan}}]{Endo:2020kro}%
  \BibitemOpen
  \bibfield  {author} {\bibinfo {author} {\bibfnamefont {S.}~\bibnamefont {Endo}}, \bibinfo {author} {\bibfnamefont {Z.}~\bibnamefont {Cai}}, \bibinfo {author} {\bibfnamefont {S.~C.}\ \bibnamefont {Benjamin}}, \ and\ \bibinfo {author} {\bibfnamefont {X.}~\bibnamefont {Yuan}},\ }\href {\doibase 10.7566/JPSJ.90.032001} {\bibfield  {journal} {\bibinfo  {journal} {J. Phys. Soc. Jap.}\ }\textbf {\bibinfo {volume} {90}},\ \bibinfo {pages} {032001} (\bibinfo {year} {2021})},\ \Eprint {http://arxiv.org/abs/2011.01382}{arXiv:2011.01382 [quant-ph]}\BibitemShut {NoStop}%
\bibitem [{\citenamefont {Koczor}(2021)}]{Koczor:2020wrs}%
  \BibitemOpen
  \bibfield  {author} {\bibinfo {author} {\bibfnamefont {B.}~\bibnamefont {Koczor}},\ }\href {\doibase 10.1103/PhysRevX.11.031057} {\bibfield  {journal} {\bibinfo  {journal} {Phys. Rev. X}\ }\textbf {\bibinfo {volume} {11}},\ \bibinfo {pages} {031057} (\bibinfo {year} {2021})},\ \Eprint {http://arxiv.org/abs/2011.05942}{arXiv:2011.05942 [quant-ph]}\BibitemShut {NoStop}%
\bibitem [{\citenamefont {Takagi}\ \emph {et~al.}(2022)\citenamefont {Takagi}, \citenamefont {Endo}, \citenamefont {Minagawa},\ and\ \citenamefont {Gu}}]{Takagi:2021nnm}%
  \BibitemOpen
  \bibfield  {author} {\bibinfo {author} {\bibfnamefont {R.}~\bibnamefont {Takagi}}, \bibinfo {author} {\bibfnamefont {S.}~\bibnamefont {Endo}}, \bibinfo {author} {\bibfnamefont {S.}~\bibnamefont {Minagawa}}, \ and\ \bibinfo {author} {\bibfnamefont {M.}~\bibnamefont {Gu}},\ }\href {\doibase 10.1038/s41534-022-00618-z} {\bibfield  {journal} {\bibinfo  {journal} {npj Quantum Inf.}\ }\textbf {\bibinfo {volume} {8}},\ \bibinfo {pages} {114} (\bibinfo {year} {2022})},\ \Eprint {http://arxiv.org/abs/2109.04457}{arXiv:2109.04457 [quant-ph]}\BibitemShut {NoStop}%
\bibitem [{\citenamefont {Peruzzo}\ \emph {et~al.}(2014)\citenamefont {Peruzzo}, \citenamefont {McClean}, \citenamefont {Shadbolt}, \citenamefont {Yung}, \citenamefont {Zhou}, \citenamefont {Love}, \citenamefont {Aspuru-Guzik},\ and\ \citenamefont {O'Brien}}]{Peruzzo:2013bzg}%
  \BibitemOpen
  \bibfield  {author} {\bibinfo {author} {\bibfnamefont {A.}~\bibnamefont {Peruzzo}}, \bibinfo {author} {\bibfnamefont {J.}~\bibnamefont {McClean}}, \bibinfo {author} {\bibfnamefont {P.}~\bibnamefont {Shadbolt}}, \bibinfo {author} {\bibfnamefont {M.-H.}\ \bibnamefont {Yung}}, \bibinfo {author} {\bibfnamefont {X.-Q.}\ \bibnamefont {Zhou}}, \bibinfo {author} {\bibfnamefont {P.~J.}\ \bibnamefont {Love}}, \bibinfo {author} {\bibfnamefont {A.}~\bibnamefont {Aspuru-Guzik}}, \ and\ \bibinfo {author} {\bibfnamefont {J.~L.}\ \bibnamefont {O'Brien}},\ }\href {\doibase 10.1038/ncomms5213} {\bibfield  {journal} {\bibinfo  {journal} {Nature Commun.}\ }\textbf {\bibinfo {volume} {5}},\ \bibinfo {pages} {4213} (\bibinfo {year} {2014})},\ \Eprint {http://arxiv.org/abs/1304.3061}{arXiv:1304.3061 [quant-ph]}\BibitemShut {NoStop}%
\bibitem [{\citenamefont {Kandala}\ \emph {et~al.}(2017)\citenamefont {Kandala}, \citenamefont {Mezzacapo}, \citenamefont {Temme}, \citenamefont {Takita}, \citenamefont {Brink}, \citenamefont {Chow},\ and\ \citenamefont {Gambetta}}]{Kandala:2017vok}%
  \BibitemOpen
  \bibfield  {author} {\bibinfo {author} {\bibfnamefont {A.}~\bibnamefont {Kandala}}, \bibinfo {author} {\bibfnamefont {A.}~\bibnamefont {Mezzacapo}}, \bibinfo {author} {\bibfnamefont {K.}~\bibnamefont {Temme}}, \bibinfo {author} {\bibfnamefont {M.}~\bibnamefont {Takita}}, \bibinfo {author} {\bibfnamefont {M.}~\bibnamefont {Brink}}, \bibinfo {author} {\bibfnamefont {J.~M.}\ \bibnamefont {Chow}}, \ and\ \bibinfo {author} {\bibfnamefont {J.~M.}\ \bibnamefont {Gambetta}},\ }\href {\doibase 10.1038/nature23879} {\bibfield  {journal} {\bibinfo  {journal} {Nature}\ }\textbf {\bibinfo {volume} {549}},\ \bibinfo {pages} {242} (\bibinfo {year} {2017})},\ \Eprint {http://arxiv.org/abs/1704.05018}{arXiv:1704.05018 [quant-ph]}\BibitemShut {NoStop}%
\bibitem [{\citenamefont {Yuan}\ \emph {et~al.}(2019)\citenamefont {Yuan}, \citenamefont {Endo}, \citenamefont {Zhao}, \citenamefont {Li},\ and\ \citenamefont {Benjamin}}]{Yuan:2018jdl}%
  \BibitemOpen
  \bibfield  {author} {\bibinfo {author} {\bibfnamefont {X.}~\bibnamefont {Yuan}}, \bibinfo {author} {\bibfnamefont {S.}~\bibnamefont {Endo}}, \bibinfo {author} {\bibfnamefont {Q.}~\bibnamefont {Zhao}}, \bibinfo {author} {\bibfnamefont {Y.}~\bibnamefont {Li}}, \ and\ \bibinfo {author} {\bibfnamefont {S.~C.}\ \bibnamefont {Benjamin}},\ }\href {\doibase 10.22331/q-2019-10-07-191} {\bibfield  {journal} {\bibinfo  {journal} {Quantum}\ }\textbf {\bibinfo {volume} {3}},\ \bibinfo {pages} {191} (\bibinfo {year} {2019})},\ \Eprint {http://arxiv.org/abs/1812.08767}{arXiv:1812.08767 [quant-ph]}\BibitemShut {NoStop}%
\bibitem [{\citenamefont {Endo}\ \emph {et~al.}(2020)\citenamefont {Endo}, \citenamefont {Sun}, \citenamefont {Li}, \citenamefont {Benjamin},\ and\ \citenamefont {Yuan}}]{Endo:2020sxk}%
  \BibitemOpen
  \bibfield  {author} {\bibinfo {author} {\bibfnamefont {S.}~\bibnamefont {Endo}}, \bibinfo {author} {\bibfnamefont {J.}~\bibnamefont {Sun}}, \bibinfo {author} {\bibfnamefont {Y.}~\bibnamefont {Li}}, \bibinfo {author} {\bibfnamefont {S.~C.}\ \bibnamefont {Benjamin}}, \ and\ \bibinfo {author} {\bibfnamefont {X.}~\bibnamefont {Yuan}},\ }\href {\doibase 10.1103/PhysRevLett.125.010501} {\bibfield  {journal} {\bibinfo  {journal} {Phys. Rev. Lett.}\ }\textbf {\bibinfo {volume} {125}},\ \bibinfo {pages} {010501} (\bibinfo {year} {2020})},\ \Eprint {http://arxiv.org/abs/1812.08778}{arXiv:1812.08778 [quant-ph]}\BibitemShut {NoStop}%
\bibitem [{Note2()}]{Note2}%
  \BibitemOpen
  \bibinfo {note} {We performed a preliminary test with the nearest-neighbor Ising model Hamiltonian as $H_d$, but the QCL training result was not as good.}\BibitemShut {Stop}%
\bibitem [{\citenamefont {Haldane}(1988)}]{Haldane:1987gg}%
  \BibitemOpen
  \bibfield  {author} {\bibinfo {author} {\bibfnamefont {F.~D.~M.}\ \bibnamefont {Haldane}},\ }\href {\doibase 10.1103/PhysRevLett.60.635} {\bibfield  {journal} {\bibinfo  {journal} {Phys. Rev. Lett.}\ }\textbf {\bibinfo {volume} {60}},\ \bibinfo {pages} {635} (\bibinfo {year} {1988})}\BibitemShut {NoStop}%
\bibitem [{\citenamefont {Sriram~Shastry}(1988)}]{SriramShastry:1987wdh}%
  \BibitemOpen
  \bibfield  {author} {\bibinfo {author} {\bibfnamefont {B.}~\bibnamefont {Sriram~Shastry}},\ }\href {\doibase 10.1103/PhysRevLett.60.639} {\bibfield  {journal} {\bibinfo  {journal} {Phys. Rev. Lett.}\ }\textbf {\bibinfo {volume} {60}},\ \bibinfo {pages} {639} (\bibinfo {year} {1988})}\BibitemShut {NoStop}%
\end{thebibliography}%
\bibliographystyle{apsrev4-2}

\end{document}